# DC Electrical Degradation of YSZ: Voltage Controlled Electrical Metallization of A Fast Ion Conducting Insulator

Ana Alvarez[1], Yanhao Dong[2], I-Wei Chen[1*]

[1] *Department of Material Science and Engineering, University of Pennsylvania, Philadelphia, PA 19104, USA*

[2] *Department of Nuclear Science and Engineering, Massachusetts Institute of Technology, Cambridge, MA 02139, USA*

**Abstract**

DC electrical degradation as a form of dielectric and resistance breakdown is a common phenomenon in thin-film devices including resistance-switching memory. To obtain design data and to probe the degradation mechanism, highly accelerated lifetime tests (HALT) are often conducted at higher temperatures with thicker samples. While the mechanism is well established in semiconducting oxides such as perovskite titanates, it is not in stabilized zirconia and other fast oxygen-ion conductors that have little electronic conductivity. Here we model the mechanism by an oxygen-driven, transport-limited, metal-insulator transition, which finds support in rich experimental observations—including *in situ* videos and variable temperature studies—of yttria-stabilized zirconia. They are contrasted with the findings in semiconducting titanates and resistance memory, and provide new insight into ceramic processing with extremely rapid heating and cooling such as flash sintering and melt processing.

*Corresponding Author Information

**Tel: +1-215-898-5163; Fax: +1-215-573-2128**

***E-mail address*: iweichen@seas.upenn.edu (I-Wei Chen)**

***Postal address*:** Department of Materials Science and Engineering, University of Pennsylvania, LRSM Building, Room 424, 3231 Walnut St., Philadelphia, PA 19104-6272

## INTRODUCTION

Resistance degradation under a sustained DC voltage stress is common in dielectrics and even more prevalent in thinner devices that increasingly appear in newer applications. [1-3] Highly accelerated lifetime tests (HALT) conducted at high voltages and temperatures provide a means to assess the degradation kinetics and mechanisms. [4] Such study has been undertaken for perovskite-type titanates as reported and summarized by Waser *et al*. almost 30 years ago. [1-3] But new interest has recently arisen from very similar findings in the forming step of resistance-switching memory [5-9] and the stressing step of flash sintering [10]. Both involving a severe resistance decrease after the dielectric is loaded by a DC voltage for some time. Remarkably, seemingly similar degradation phenomena have been seen in a large number of fundamentally different dielectrics, for example titanates that are semiconductors vs. zirconia that are fast ion conductors. Such phenomenological similarity must be superficial, for the increasing leakage current in titanates comes from electrons and holes with almost no contribution from the slow-moving oxygen ions. Indeed, in the study of Waser *et al*. and subsequent work [1-3,11-13], oxygen ions were assumed to be entirely blocked by the electrodes, and it is the gradual, voltage-motivated redistribution of oxygen (or oxygen vacancies) between the electrodes that controls the kinetics of leakage-current increase. Clearly, such mechanism is not applicable to fast oxygen ion conductors, in which oxygen-ion current completely dominates over electronic current [14], and which are often used to build solid-electrolyte devices with relatively robust oxygen conductance at electrodes. Recognizing the knowledge gap, this paper hopes to provide a comprehensive theoretical and experimental picture of HALT for a prototypical fast ion conductor, the cubic phase of 8 mol% yttria-stabilized zirconia, called 8YSZ, or simply YSZ.

As a leading solid electrolyte for sensing and high-temperature energy storage/generation applications [15-17], 8YSZ must avoid fading of galvanic voltage and provide environment-insensitive conductivity. This is achieved by minimizing electronic conductivity and maintaining a large, stable ionic conductivity with the aid of $Y^{3+}$ doping, which provides a fixed number of fast-moving $O^{2-}$ or its point defect $V_O^{\cdot\cdot}$ over a very wide range of electrochemical conditions, known as the electrolytic range. In 8YSZ, the range spans over 40 decades of oxygen partial pressure as shown in **Fig. 1**. [14] In fact, since the activation energy of electron and hole conduction is several times that of oxygen in 8YSZ [18,19], the range is even wider at lower temperatures. On the other hand, at lower temperatures, 8YSZ is also an excellent insulator; indeed, its "first cousin" $HfO_2$ dielectric (known as the gate oxide in integrated circuits and having a thickness of just a few atomic layers) can withstand a huge voltage stress while storing capacitive charge without leakage. These features are not shared by titanates, whose leakage current sensitively depends on the atmosphere as do their concentrations of oxygen vacancies, electrons and holes. [1-3,11-13] This comparison again makes clear that YSZ follows a totally different degradation mechanism from titanates.

To obtain some clue of the new mechanism, we begin by questioning how resistance degradation is possible at all if a constant ionic conductivity is indeed maintained in 8YSZ over a very wide electrolytic range. The answer must be: the fast ion conductor has broken down, and the product no longer possesses the same constant ionic conductivity. So the central theoretical issue becomes the thermodynamics and kinetics of this break-down process, which we will ascribe to a first-order insulator-to-metal transition with a sharp phase-interface associated with a definite chemical potential. In the zirconia literature, Janek and Korte suggested that this interface is demarcated by electro-coloration, which separates "black" zirconia from "white" or colorless transparent zirconia. [20] While the former is indeed more conductive than the latter [21], it is by no means always metallic; in fact, according to our study, it may only be modestly more conductive. Since blackness cannot be taken as a marker of a metal and is a poorly quantifiable color anyway, a more direct experimental evidence of the metallic phase is needed. This was obtained despite the difficulty posed by the fact that YSZ is a fast oxygen conductor and, as will become clear later, highly defective zirconia suboxides are precarious metals.

## THEORY

### Phase equilibrium between zirconia and metallic suboxide

We first ignore Y since cations being immobile at low temperatures must have identical composition in different phases during phase transition. We consider one insulating dioxide $ZrO_{2-\varepsilon}$, containing some $O^{2-}$ vacancies, as the parent phase and one metallic suboxide as the product phase. Without loss of generality, we let the metallic phase be $ZrO_{1+\delta}$, containing some $O^{2-}$ interstitials, although other suboxides could be considered instead (e.g., $ZrO_{1.5}$ is known to be metallic [22], and there are other suboxides close to being metallic. [23]) The three-phase equilibrium, of the above two solids and a $O_2$ gas, at a partial pressure $PO_2$, obeys the phase rule $P = C - F + 2$, having $P = 3$, $C = 2$, and one degree of freedom $F$ along the $PO_2(T)$ isotherm. Therefore, at a given temperature, three-phase equilibrium occurs at a unique $PO_2^{\mathrm{c}}$, or equivalently, a unique chemical potential $\mu_{O_2}^{\mathrm{c}}$.

We next derive $PO_2^{\mathrm{c}}$ and $\mu_{O_2}^{\mathrm{c}}$ using a simple model to gain some insight. Denoting the chemical potentials and the electrochemical potential of the $i$-th species by $\mu_i$ and $\tilde{\mu}_i$, respectively, we use the redox reaction

$$\frac{1}{2}O_2 + 2e = O^{2-} \qquad (1)$$

to prescribe the following equilibrium conditions for the potentials

$$1/2\,\mu_{O_2} + 2\mu_e = \mu_{O^{2-}} \qquad (1a)$$

$$1/2\,\mu_{O_2} + 2\tilde{\mu}_e = \tilde{\mu}_{O^{2-}} \qquad (1b)$$

In the above, $\tilde{\mu}_i \equiv \mu_i + z_i e\phi$, where $\phi$ is the electric potential, $e$ the elementary charge, and $z_i e$ the charge of the *i*-th species. In the metallic phase, $\mu_e$ is the Fermi level of electrons, which is unaffected by the redox reaction and independent of $PO_2$ because metal has a very large number of delocalized electrons. Therefore, inserting $O^{2-}$ interstitials into the suboxide alters only $\mu_{O^{2-}}$ and $\mu_{O_2}$, following $\mu_{O^{2-}} = 1/2\,\mu_{O_2} + \text{constant}$ where constant = $2\mu_e$. Third, taking into account the large amount of $Y^{3+}$ in YSZ that causes a fixed concentration of oxygen vacancies to form to satisfy charge neutrality, we see $\mu_{O^{2-}} = \text{constant}$, which also explains the wide electrolytic range found in YSZ. Fourth, applying Eq. (1a) to YSZ, we obtain $\mu_e = -1/4\,\mu_{O_2} + \text{constant}$ where $\text{constant} = 1/2\,\mu_{O^{2-}}$, meaning that with a decreasing $\mu_{O_2}$ the Fermi level in YSZ increases as evident from the increasing electron concentration in **Fig. 1**. This naturally leads to a critically low $PO_2^c$ or $\mu_{O_2}$, which is the condition when the Fermi level $\mu_e$ of YSZ reaches the Fermi level of the metallic suboxide.

In summary, in the metallic phase, $\mu_{O^{2-}}$ varies with $\mu_{O_2}$ but $\mu_e$ is fixed, and in the ionic phase, $\mu_e$ varies with $\mu_{O_2}$ but $\mu_{O^{2-}}$ is fixed. The two phases share a common, unique $\mu_{O_2}^c$ at their interface, and once the $\mu_{O_2}$ distribution on two sides of the interface is known, the distributions of $\mu_{O^{2-}}$ and $\mu_e$ are too known from Eq. (1a). Note that the chemical potentials of cations are not needed because they are immobile at the typical temperatures of HALT and device operations. Also, since $\mu_{\mathrm{O}^{2-}}$ is fixed in the ionic phase, the variation of the $1/2\,\mu_{\mathrm{O}_2}$ distribution and the $2\mu_e$ distribution, which are both highly non-linear, must mirror each other according to Eq. (1a). Further discussion of the $\mu_{\mathrm{O}_2}$ distribution in YSZ is not needed for the following development and can be found elsewhere. [24,25]

## Steady state solution and the critical voltage for the formation of metallic phase

Consider a one-dimensional slab of YSZ in air having a voltage $V$ applied across its length $L$, driving a (negative) current to flow from right (the anode) to left (the cathode). Below we show that no metallic suboxide should form up to a critical voltage $V^{\mathrm{c}}$ in the steady state. This is ascertained by providing a solution of the $\mu_{\mathrm{O}_2}$ distribution that (a) lies above $\mu_{\mathrm{O}_2}^{\mathrm{c}}$ everywhere, and (b) allows the same oxygen flux to flow throughout the length and the two electrodes. This includes $\mathrm{O}^{2-}$ entering from the cathode side and exiting from the anode side, but no oxygen is allowed to enter or exist from the lateral sides. Details of the solution procedure is provided in the online version of this manuscript. [26]

We now model the kinetics at the air-electrode interfaces assuming it is interface-controlled. We use the simplest such kinetics, which is linear kinetics: the current density is proportional to the driving force, which is the difference of $\mu_{\mathrm{O}_2}$ across the interface, as schematically shown in **Fig. 2a**. Specifically, the cathode current density is given by $j^{\mathrm{cathode}} = k'(\mu'_{\mathrm{O}_2} - \mu_{\mathrm{O}_2}^{\mathrm{air}})$, where $k'$ is a rate constant and $\mu'_{\mathrm{O}_2}$ is the cathode's $\mu_{\mathrm{O}_2}$, lower than that of air, $\mu_{\mathrm{O}_2}^{\mathrm{air}}$. Likewise, the anode current density is $j^{\text{anode}} = -k''(\mu''_{O_2} - \mu^{\text{air}}_{O_2})$, where $k''$ is another rate constant and $\mu''_{O_2}$ is the anode's $\mu_{O_2}$, higher than $\mu^{\text{air}}_{O_2}$. Since the current density, hence the ability to admit $O^{2-}$ at the cathode, reaches the upper limit when the metallic phase forms there, at $\mu'_{O_2} = \mu^{c}_{O_2}$, there is a maximum current density allowed to flow through YSZ at the steady state. Anode is immaterial because there is no similar limit there for releasing $O_2$.

We next consider the ion transport in YSZ, which is a fast $O^{2-}$ conductor. Since $\mu_{O^{2-}}$ is constant in YSZ, $O^{2-}$'s movement is driven by the electrical field only, which must be constant at the steady state. Therefore, the current density in YSZ is $j^{\text{YSZ}} = -\sigma_{O^{2-}}\Delta\phi / L$, where $\sigma_{O^{2-}}$ is the $O^{2-}$ conductivity and $\Delta\phi$ is the electric potential difference between the anode and the cathode.

We now use flux and potential continuity to obtain the steady state solution. For any prescribed $\Delta\phi$, $j^{\text{YSZ}}$ is given, and the boundary $\mu_{O_2}$ at the electrodes is obtained from $j^{\text{YSZ}} = j^{\text{cathode}} = j^{\text{anode}}$. However, since the current in the slab is actually controlled by the voltage *V*, we should relate $\Delta\phi$ to *V*, which is the difference between the Fermi levels—namely the electrochemical potentials of electron—of the cathode and the anode. From Eq. (1b), the difference is $\Delta\tilde{\mu}_e = -\Delta\phi - (\mu''_{O_2} - \mu'_{O_2})/4e$. (In writing the above, we recognized that $\mu_{O^{2-}}$ is constant.) So,

$$V = -\Delta\tilde{\mu}_e = \Delta\phi + (\mu''_{O_2} - \mu'_{O_2})/4e \qquad (2)$$

Expressing the current density $j = j^{\text{YSZ}} = j^{\text{cathode}} = j^{\text{anode}}$, which is negative, in terms of *V*, we have

$$V = (|j| L/\sigma_{O^{2-}})(1 + \sigma_{O^{2-}}/4eLk' + \sigma_{O^{2-}}/4eLk'') \qquad (3)$$

Here, keeping only the first term in the second bracket on the right gives just the Ohm's law, and adding the equivalent resistance of electrodes relative to the resistance of YSZ of a unit area *A*, $L/\sigma_{O^{2-}}$, gives the other two terms, $1/4eLk'A$ for $R^{\mathrm{cathode}}$ and $1/4eLk''A$ for $R^{\mathrm{anode}}$.

The above form of steady state solution is well known in transport problems with interface-limited kinetics. If $k'$ and $k''$ are constant, they will each add a constant resistance to the total resistance, which is still Ohmic, meaning that the resistance is independent of *V*. This breaks down if $k'$ and $k''$ themselves depend on $\mu'_{O_2} - \mu^{\mathrm{air}}_{O_2}$ and $\mu''_{O_2} - \mu^{\mathrm{air}}_{O_2}$, as in the so-called Butler-Volmer kinetics. However, when the electrode resistance is relatively small, which happens when $k'L$ and $k''L$ are large, then the device may still appear relatively Ohmic with essentially the YSZ resistance despite having "imperfect" electrodes.

The steady state solution obtains until a critical voltage $V^{\mathrm{c}}$, which is set by replacing $|j|$ by $|j'_{\max}| = k'(\mu^{\mathrm{air}}_{O_2} - \mu^{\mathrm{c}}_{O_2})$, the maximum current density set by the cathode,

$$V^{\mathrm{c}} = (k'(\mu^{\mathrm{air}}_{O_2} - \mu^{\mathrm{c}}_{O_2})L/\sigma_{O^{2-}})(1 + \sigma_{O^{2-}}/4eLk' + \sigma_{O^{2-}}/4eLk'') \qquad (4)$$

It increases with $k'$ because of a larger $j'_{\max}$ allowed, and decreases with $k''$ because of a lesser anode resistance allowed. So the critical voltage is very sensitive to the electrode kinetics and conditions (electrode roughness, porosity, microstructure, composition, etc.). In this work, we treat it as an empirical property to be determined experimentally. Note that $k' = 0$ is mathematically and physically a singular case. Since the blocking cathode allows no current, $V^{\mathrm{c}} = 0$ necessarily. At $V = 0$, there is the trivial solution of $\mu_{O_2} = \mu'_{O_2} = \mu''_{O_2} = \mu^{air}_{O_2}$, but at any $V > 0$, it has no steady-state solution because immediately $\mu'_{O_2} = \mu^{c}_{O_2}$ and the metallic phase forms at the anode.

**Growth of the metallic phase and resistance degradation**

The situation of $V > V^c$ is depicted in **Fig. 2b**, which shows a metallic phase extending from the cathode over a length of $(1-f)L$, and YSZ extending from the anode over a length of $fL$. This happens when less oxygen enters the cathode than leaves the anode, which depletes the oxygen content in the sample. Since YSZ must have a fixed oxygen content as emphasized before, the depletion will cause a part of it—the part next to the cathode—to convert to the metallic suboxide. As time $t$ increases, $f$ decreases, and the phase boundary moves to the right at a time-dependent (or position/*f*-dependent) velocity $v$.

To solve $v$ and $f$, we again consider current and potential continuity in the sample, which now has a metallic phase in addition to YSZ and electrodes. We seek the quasi-steady-state solution, which has (a) a constant current flowing through the cathode and the metallic phase and (b) another constant current flowing through YSZ and the anode. However, because the oxygen content is lower in the metallic phase than in YSZ, the current (a) is smaller than the current (b) with the difference made up by the interface movement, which is equivalent to a convection current.

Since a metal cannot support any electric field, the (interstitial) $O^{2-}$ flux must come from a gradient of $\mu_{O^{2-}}$, which is the same as a gradient of $\mu_{O_2}$ because $\mu_{O^{2-}} = 1/2\,\mu_{O_2} + \text{constant}$ where constant = $2\mu_e$. (This does not violate the phase rule, for the $\mu_{O_2}$ variation is in the single-phase region, not at the two-phase interface.) Thus, the $O^{2-}$ current density is $j^{metal} = ec(D/k_B T)\partial\mu_{O_2}/\partial x$, where $D$ is the self-diffusivity of $O^{2-}$ interstitial, $k_B T$ has its usual meaning, $x$ is the distance from the cathode, and $c$ is the concentration of interstitials, which may be plausibly assumed to be constant because the suboxide is likely to be defect-rich and a fast oxygen conductor itself. Therefore, at the quasi-steady state, $\partial\mu_{O_2}/\partial x$ must be constant and may be replaced by $\Delta\mu_{O_2}/(1-f)L = (\mu^{c}_{O_2} - \mu'_{O_2})/(1-f)L$. Meanwhile, as the metallic phase grows, to support the same applied voltage the electric field must increase in YSZ, which increases the current density $j^{YSZ} = -\sigma_{O^{2-}}\Delta\phi / fL$. Here, $\Delta\phi$ is related to $V$ by

$$V = \Delta\phi + (\mu''_{O_2} - \mu^{c}_{O_2})/4e \qquad (5)$$

which is different from Eq. (2) because the Fermi level in the metal is constant. Lastly, the mismatch between $j^{metal}$ and $j^{YSZ}$ determines the velocity of the interface, across which there is an excess oxygen concentration $\Delta C$. (YSZ, being $0.92ZrO_2$+$0.08Y_2O_3$, has about 1 more oxygen every cation than the suboxide, although the exact value of $\Delta C$ depends on the valence of Y in the suboxide, some conducting ones may offer a much smaller $\Delta C$. [22-23]) That is, the interface movement help ensure $O^{2-}$ continuity from the metallic phase to YSZ.

The above consideration of current and potential continuity gives the following solution of $v$

$$v(2e\Delta C) = (\mu^{c}_{O_2} - \mu^{air}_{O_2})/\left[(1/k') + (1-f)L/k^{metal}\right] + \left[(\mu^{c}_{O_2} - \mu^{air}_{O_2}) + 4eV\right]/\left[1/k'' + fL4e/\sigma_{O^{2-}}\right] \quad (6)$$

where $k^{metal}$ is a shorthand for $ec(D/k_B T)$. It can be shown that, at $f=1$, $v=0$

when $V = V^{c}$, which reduces to the steady-state solution obtained in Eq. (7). When $V > V^{c}$, $v$ is always positive meaning $f$ must decrease with time, starting from $t = 0$, given by $dt = -L\,df/v$. Using $v(f,V)$ of Eq. (6) and integrating, we obtain $f(t)$ and the lifetime $\tau$, the time it takes for $f$ to go from 1 to 0. In the above, only mass continuity for $O^{2-}$ was considered. But in the cathode and the metallic suboxide there is also an injected electronic current density of a magnitude $v\Delta C$, which does not require any voltage because it passes through a metal. The injected electronic current is completely used up to convert $Zr^{4+}$ in YSZ to $Zr^{2+}$ in ZrO at the metal-YSZ interface, so it does not continue into YSZ. The solution is now complete.

The solution $df/dt$ can be cast into a dimensionless form in terms of a characteristic voltage, $V^{*} = (\mu_{O_2}^{\text{air}} - \mu_{O_2}^{\text{c}})/4e$, and a characteristic time, $t^{*} = (LA2e\Delta C)R^{YSZ0}/V^{*}$, where $R^{YSZ0} = L/A\sigma_{O^{2-}}$ is the resistance of undegraded YSZ of a cross section A, and $LA2e\Delta C$ is the total amount of charge required to convert $ZrO_2$ to ZrO. The dimensionless solution is

$$-df/d(t/t^{*}) = -1/\left(R^{\text{cathode}}/R^{\text{YSZ0}} + (1-f)\,R^{\text{metal}}/R^{\text{YSZ0}}\right) + \left(V/V^{*} - 1\right)/\left(R^{\text{anode}}/R^{\text{YSZ0}} + f\right) \quad (7)$$

In the above, $t/t^{*}$ is the reduced time, $V/V^{*}$ is the reduced voltage, $1/4eLk'A$ is $R^{\text{cathode}}$, $1/4eLk''A$ for $R^{\text{anode}}$, and $R^{\text{metal}} = L/4ek^{\text{metal}}A$ is the (diffusion) resistance of ZrO of the same dimension as the original YSZ.

This equation has the form of $-df/d(t/t^{*}) = -1/\left(a + (1-f)b\right) + \left(V/V^{*} - 1\right)/\left(c + f\right)$, so the critical voltage for getting a nonnegative $df/dt$ is $V^{c}/V^{*} = 1 + c/a + 1/a$, which reduces to Eq. (4), as expected. From this simple form, we can immediately identify two limiting cases: (i) perfect cathode ($R^{\text{cathode}} = 0$) giving $a = 0$, which has no resistance degradation; (ii) blocked cathode giving $1/a = 0$, which as mentioned before is a special case and the degradation starts at zero voltage. For any other value of *a*, at $V/V^*$ greater than $1 + c/a + 1/a$, we can integrate Eq. (7) to obtain $f(t/t^*)$, which is the YSZ resistance at time *t* divided by its initial value $R^{\mathrm{YSZ0}}$. It should be very close to $R(t)/R(0) = r$, the reduced slab resistance, which is the slab resistance at time *t*, $R(t)$, divided by its initial value $R(0)$. Below, we will simplify the notation by writing $R(t)$ as $R$, $R(0)$ as $R_{\mathrm{o}}$, and $r = R/R_{\mathrm{o}}$.

We have plotted the numerical solutions of $r$ in **Fig. 3** for a number of cases. They clearly show a generic shape of accelerated degradation (**Fig. 3a**), a very pronounced effect of increasing *c* (namely increasing $R^{\mathrm{anode}}$) on slowing down degradation, and a lack of effect of changing *b* (namely changing $R^{\mathrm{metal}}$). Although not shown here, we also verified that the degradation kinetics is relatively insensitive to *a*, as long as it does not fall to the two extremes mentioned above. The above finding is further confirmed by the plots of "lifetime" $\tau$ in **Fig. 3b**. This time increases with *c* but is very insensitive to *a* and *b*. Not surprisingly, it always rapidly decreases with *V*.

## EXPERIMENTS

### Experimental procedure

We performed over 330 HALT on 8YSZ single crystals (MTI Corp., Richmond, CA) and polycrystals (with a grain size 1.9 μm and a relative density better than 96%), tested from 200°C to 500°C. In most cases, test bars of two lengths, 5 and 10 mm, with a width of 1.6 mm and a thickness of 0.5 mm, and with two electrodes covering the two ends on the top face (**Fig. 2**), were used. In the remaining tests, 0.5 mm thick disks with electrodes on two opposite flats were used. These electrodes were made by applying a pure Pt paste (Fuel Cell Materials, Lewis Center, OH) to the sample surface, followed by annealing at 1000ºC for 2 h. Most tests were conducted in air, from 200-440ºC, in a furnace or on a hot plate. The latter setting facilitated *in situ* observations using a video camera, sometimes aided by additional lenses and/or a thermal camera. Hot-plate tests were also performed to observe oxygen bubble evolution, by immersing the sample in a silicone fluid, which has a high flash point (314-359ºC depending on the composition) and excellent oxygen solubility and permeability. Generally speaking, the degradation kinetics increased with temperature and electric field, and all the results and observations were broadly consistent with each other, so only a summary is provided below. Additional video clips on oxygen bubble evolution, black band propagation and recession, and sample flashing at the end of the degradation test are also provided in the online version of this manuscript. [26] Finally, similar to what Waser et al. had reported [1-3], we found considerable scatter of the degradation time $\tau$ and the initial resistance $R_o$, and within the scatter no systematic difference between polycrystals and single crystals was apparent.

## Generic degradation curves

Degradation curves shown in **Fig. 4** have a generic shape with three stages. In stage I, the resistance degradation accelerates and the $r(t)$ curve has a concave downward curvature similar to those in **Fig. 3a**. This stage becomes very short when degradation is too fast. In stage II, deceleration takes hold and the $r(t)$ curve has a concave upward curvature. In stage III, acceleration resumes and the $r(t)$ curve has a concave downward curvature falling toward $r = 0$. To arrest the final breakdown, we often set a current compliance of either 0.2 A or 0.02 A, and the same compliance was used at different temperatures despite the initial resistance $R_o$ is strongly temperature dependent, e.g., ~10 MΩ at 325°C and ~400 MΩ at 250°C. With the latter setting and an applied voltage of 200 V, the resistance at 325-350°C typically stayed on the order of $10^5$ Ω during stage II, then descended rapidly spending a brief time on the $10^4$ Ω range in stage III before reaching a resistance of 2 – 5 kΩ when the compliance was triggered, upon which the voltage between the electrodes immediately dropped by some 40-60%.

**Blackening does not equal metallic conduction**

Electro-coloration occurs in YSZ because of electron association with point defects, forming color centers, and very evident blackening can be produced without destroying the crystal structure (up to 3 mol% extra oxygen vacancies). [20,27,28] During HALT, we always observed various degrees of blackening initiating from the cathode and propagating toward the anode. Often, the black region was first confined to one or two fingers (**Fig. 5a**), but by the end of the test the entire YSZ was invariably black. The blackened region is not necessarily metallic. This is evident from **Fig. 5a-b** in which the tips of one or more blackened fingers have already reached the anode, yet the overall resistance degradation is <30% in the case of **Fig. 5a** and <20% in the case of **Fig. 5b**.

If the blackened region were indeed metallic, then it would have shorted the two electrodes and the resistance would have dropped drastically, which is not seen in **Fig. 5a-b**.

Nevertheless, in stage III, most of the blackened region is likely metallic. If we assume the metallic region also begins in a finger-like manner before gradually widening and branching out, then we may regard the former stage as the growth stage and the latter stage the "coarsening" stage. Electrically, the growth stage is like extending a metallic element from cathode to anode, which results in a short circuit and a drastic resistance drop, while the coarsening stage is like widening the cross section of the short circuit, which causes a gradual and progressively less severe decrease of the already small resistance. This explains the opposite concavity of Stage I and II, the former corresponding to the growth stage, the latter the coarsening stage.

Fingering is a well-known instability in first-order phase transitions. [20,29,30] Its appearance means that our simple theory that assumes one-dimensional growth is not strictly valid because the actual electric field, current, and diffusional fluxes are all concentrated at the finger-tip, which is a three-dimensional problem. [29] Therefore, despite the same physics, the actual growth of the finger should be much faster than predicted by our theory.

**Direct evidence of metallic phase**

Direct evidence of a metallic region growing from the cathode was provided by the following "field-cool" experiment. In field-cool, the test was terminated but the voltage was kept while the sample was allowed to cool. The voltage was later removed at low temperature and the electrodes were next removed to examine the areas underneath. As shown in **Fig. 6**, the area beneath the anode is black but not beneath the cathode. This asymmetry is explained by forming a metallic phase next to the cathode but not the anode. Specifically, if a metallic "sheath" forms to surround the leading edge of the cathode, it will shield the electric field from the rear side of the cathode and will, in effect, move the electrode to the front edge of the metallic region. So there is no blackening of the entire cathode region. In contrast, lacking any metallic phase next to it, the entire anode serves as a collector plate receiving electric current and field that emanate from the front of the metallic region; this produces blackening of the entire anode region.

When "zero-field-cool" was used instead, we found no black region underneath either electrodes. This suggests that it is rather easy to destroy the color centers, at least near the anode. This is not surprising. The near-anode region is mostly ionic YSZ having a nearly constant concentration of oxygen vacancies and very few electrons/holes, so it serves as a source of oxygen vacancies and sink of electrons and holes, which promotes the dissociation of the electron-defect complexes.

Curiously, after the compliance control was triggered, we often observed some regions, side by side to the blackest band, became less black or even turned white (**Fig. 7**). At this stage, the blackest band is no doubt metallic. So it can electrically unload the nearby less metallic (or even insulating) region and become the sink of latter's electrons and holes. This is similar to the case of "zero-field cool/anneal", which explains why the white region can reappear at such late stage of degradation.

**Electrode resistance**

Because only the leading edge of cathode is electrically active whereas the entire area of anode is, one expects $R^{\text{cathode}}$ much larger than $R^{\text{anode}}$. The large anode area available also makes it more likely to behave as a perfect electrode, while the opposite holds for the cathode. This was confirmed by two experiments. First, we performed HALT inside a silicone fluid, and directly observed bubble formation from the anode (**Fig. 8a**). As the resistance degraded over time and the current increased, so were the number of bubbles formed (**Fig. 8b**). Oxygen bubbling was successfully observed over the temperature range of 235-290°C, and it provided an unequivocal evidence of an efficient anode capable of emitting oxygen gas, unlike the blocked electrodes on titanates at comparable HALT temperatures. [1,2]

Second, we performed three sets of HALT with electrodes of different "goodness": (a) both electrodes were Pt-plated areas, (b) anode was fully Pt-plated but cathode was a single Pt wire-loop (see schematic in **Fig. 9**), and (c) cathode was fully Pt-plated but anode was a single Pt wire-loop. In **Fig. 9**, we found (c) alone slows down the degradation significantly, whereas (a) and (b) are almost indistinguishable. This may be understood as follows. First, since only the leading edge of a fully plated cathode is electrically effective whereas its trailing side is electrically shielded, a fully plated cathode is not very different from a single wire-loop, hence relatively little difference between (a) and (b). Second, since the full area of the plated anode in (a) is a collector plate, it has a much lower $R^{\text{anode}}$ than (c). Therefore, $(a) = (b)$ but (c) is different. Lastly, just as predicted by our theory, (c) delays degradation: this is the case of small $k''$ hence a large $R^{\text{anode}}$, which gives the curves of large *c* (i.e., large $R^{\text{anode}}$) in **Fig. 3a-b**.

Although the cathode is in effect an "edge electrode" and has a large $R^{\text{cathode}}$, its efficiency is not zero. This is because at low voltages, the resistance of our samples is Ohmic giving no indication of the influence of an electrode overpotential. Indeed, the sample resistance agrees quite well with the literature data of $O^{2-}$ conductivity suggesting there is enough electrode efficiency overall to support a normal $O^{2-}$ flow. Therefore, metallic phase only forms above a critical voltage when the cathode $\mu'_{O_2}$ has dropped to $\mu^{c}_{O_2}$.

**Critical voltage**

To determine the critical voltage, we used a single crystal and repeatedly recorded its resistance degradation curve from $r = 1$ to $r = 0.3$ at 325°C under different voltages. Between repeated runs, the voltage was reduced to 7.5 V and held for 100 s, which was enough for the resistance to recover to the same initial value $R_o$; after that a new voltage was applied. The procedure was performed in both an increasing voltage sequence and a decreasing voltage sequence, and the measured degradation rates in the two sequences were averaged and shown in **Fig. 10.** The data extrapolate to ~16 V at zero degradation, which is taken to be the critical voltage $V^{c}$.

## Degradation curves have invariant features despite large scatter

Our theory suggests that $r$ againsť $t/R_{\mathrm{o}}$ should be relatively invariant. This is because $t^* \sim R^{\mathrm{YSZ0}}$, and if $\mu_{\mathrm{O_2}}^{\mathrm{air}} - \mu_{\mathrm{O_2}}^{\mathrm{c}}$ is relatively constant and $R^{\mathrm{cathode}}$, $R^{\mathrm{anode}}$, and $R^{\mathrm{metal}}$ are all small compared to $R^{\mathrm{YSZ0}}$, then Eq. (7) predicts a universal curve for $r$ against $t/R_{\mathrm{o}}$. Since $R^{\mathrm{YSZ0}}$ is a strong function of temperature, for example it is 10 MΩ at 325°C and 400 MΩ at 250°C, checking this prediction at different temperature would provide a non-trivial test. This is shown in **Fig. 11** for four sets of $r - t/R_{\mathrm{o}}$ data recorded in one single crystal from 250°C to 325°C under the same voltage of 125 V. (Between measurements at different temperatures, the crystal was unloaded to 0 V and heated to 325°C and held there for 5 min until $R_{\mathrm{o}}$ recovered to the expected initial value of an undegraded sample.) These curves vary in time by less than 3x, compared to the 40x variation of $R_{\mathrm{o}}$, thus lending some support to the idea that they are temperature insensitive.

On the other hand, even at the same test temperature and voltage, we observed a large scatter of $\tau$ in HALT, also reported for titanates. [1-3] Most of the variation appeared in the latter part of stage II and stage III, which consumes most of the lifetime as apparent from **Fig. 4**. This is illustrated in **Fig. 12** for a number of HALT at 325°C. It can be seen that the spread of the initial portion of the curves (stage I and early stage II) shown in the inset is relatively smaller than that of the latter portion (later stage II and stage III) in **Fig. 12a**. This is further quantified in **Fig. 12b**, which plots the ratio of the standard deviation (SD) of time at the same $r$ over the mean of such time. The spread reaches a minimum at $r = 0.7$ and increases very rapidly below $r = 0.4$.

(Because $t = 0$ at $r = 1$, the ratio is undefined there.)

## Resistance degradation ends in a flash

In most cases, the HALT ended in a spectacular flash as in **Fig. 13a**. The flash started from the anode side but soon the entire section between the two electrodes lit up to red hot, which could continue indefinitely under a compliance control. Because electrode protrusion is a field/current concentrator, it often initiated the flash, **Fig. 13b**. Sometimes the flash was terminated prematurely because the anode had popped off, which "quenched" the sample and left a blackened region with a white spot at the protrusion (**Fig. 13c**), indicating the color centers there had been annealed away. Samples with a long flash (under compliance) in effect experienced full annealing, so they were largely white or colorless when examined later. Some degraded samples did not end with a flash because of loss of anode adhesion during late degradation. Similar anode degradation and decohesion due to oxygen bubble formation, at the anode/YSZ interface and inside YSZ nearby, has also been reported in solid oxide electrolyzer cells, [25,31-33] which are electrically loaded in the same manner as in our HALT except for at much higher temperature.

Flash occurs because the power dissipated in the device, $V^2/R$, essentially increases without bound toward the end of HALT. Because YSZ is responsible for almost all the resistance, all the power is consumed in YSZ of a length $fL$ that is nearly proportional to $R$. Thus the power density in YSZ is proportional to $(V/R)^2$ or (current) $^2$ or $1/f^2$. As $r \sim f$ approaches zero, the power density becomes unbounded and focused at the remaining insulating YSZ “bridge” next to the anode, which leads to thermal runaway. Once the “bridge” flashes, heat is conducted through the metallic suboxide to the rest of the sample, and the sample works like a Nernst lamp of a power $V^2/R$.

**Degraded sample manifests metal-insulator transition**

To better capture the flash and to define the conducting state of severely degraded samples, we added a serial load resistance $R_\mathrm{L}$ to the test circuit to limit the maximum current to $V/R_\mathrm{L}$. We repeatedly degraded a single crystal at 350°C under 200 V until the circuit resistance reached $2R_\mathrm{L}$, i.e., the sample resistance no less than $R_\mathrm{L}$, then let the sample field-cool from 350°C to 40°C. (In the case of $R_\mathrm{L} = 51\,\mathrm{k\Omega}$, the test was terminated when the the resistance reached $210\,\mathrm{k\Omega}$.) We found flash when $R_\mathrm{L} < 7\mathrm{k\Omega}$. When $R_\mathrm{L} \geq 8\mathrm{k\Omega}$, there was no flash, and stage III as well as the total time to degrade was greatly lengthened, by 10x or more, because voltage must be shared with the load. As shown in **Fig. 14**, the resistance initially rises slowly below 350°C, which is a behavior commonly found in “bad metals” that have temperature-insensitive carrier concentrations but temperature-sensitive scattering by very high densities of defects or impurities (such as oxygen and oxygen vacancy). But this came to a sudden end at a temperature $T_\mathrm{c}$ when the resistance abruptly increased by up to 5 orders of magnitude, after that the temperature dependence steepens remarkably, which is the behavior of an insulator. We suggest this transition, some occurring at multiple $T_\mathrm{c}$ in **Fig. 14**, is a result of a strong random potential (presumably from the oxygen impurities/vacancies in the metallic suboxide) that is known to cause electron localization below $T_c$. [34] The first $T_c$ during cooling generally increases with $R_L$, suggesting a less stable metallic state if it has more defects and impurities. Interestingly, if the sample after flashing was instead cooled under an instrument-set compliance of 0.02 A (nominally corresponding to the current $V/2R_L$ expected for $R_L = 5\ k\Omega$), then it remained in the metallic state all the way down to 40ºC (**Fig. 14**, broken curve). So a current can maintain a metallic state, which was previously seen in a ruthenate. [35] These results reflect the precarious nature of the electronic state in the presence of a random potential in a highly defective suboxide that no doubt also sees some oxygen diffusion and continuous pumping away of oxygen during the initial part of cooling. Additional data plots recording resistance and temperature vs time are provided in the on-line version of this manuscript. [26]

The above results, despite the complications that YSZ is a fast oxygen conductor and the precarious metallic phase is prone to transitions to insulators during annealing or cooling, have firmly established the fact that the conducting path between the two electrodes must have become entirely metallic once the resistance has degraded enough to flash. Even if a flash did not actually occur but the degradation was brought to near completion, then on the conducting path the last remaining "bridge" next to the anode is still metallic although it easily transforms to an insulator during cooling. (Therefore, post mortem examination at room temperature, which sees only an insulating behavior, would have missed the metallic property that invariably emerges after severe degradation.) Regarding the metal-to-insulator transition, it is not clear where it happens. It could be next to the anode, next to the cathode (the $\mu_{O_2}$ is higher there, **Fig. 2b**, and white bands appeared there in **Fig. 7**), or at multiple sections (which may account for the multiple transitions seen in **Fig. 14**).

## Repeating HALT can reduce data scatter

Because fingering is an instability [29], there is much inherent variability in its morphology, subsequently inherited by coarsening and branching, which could cause a large variability in the later part of stage II. The experiment on flash and the resistance state of the degraded sample further suggests that "breaking the final bridge" next to the anode is also highly variable, depending on the local voltage and current, which could add additional variability to stage III and $\tau$. Indeed, a slight increase in $R_L$ caused $\tau$ to increase by 10x. However, we found much less variability if the same sample was repeatedly degraded under the same voltage, after rejuvenation by an intermediate annealing subsequent to each test. To further reduce the possibility of voltage shock, we also ramped up the voltage gradually, over 600 s, in the beginning of each test. In **Fig. 15**, the results of 66 tests from 15 repeatedly tested samples are summarized in terms of $R_o$ and $\tau$, counting the time from the end of the initial voltage ramp (at 600 s) to the start of the 0.02 A compliance. Their standard deviation appears to narrow with repeated tests, which suggests that the large scatter previously seen in **Fig. 12a-b** was mostly due to sample to sample variation because of differences in defect distribution, $R_o$, electrode condition, surface condition, etc. In addition, in the same sample with the same current/field concentrators and especially after repeated testing, it is possible that fingering may initiate at the same locations and develop into a similar morphology, which sets the stage for relatively similar coarsening and final flashing.

Remarkably, we found the convergence achieved by repeated testing and annealing was destroyed after a new test condition was adopted, for example testing at a different temperature. Therefore, once the converging path is upset, variability resumes. This observation strongly suggests that variability is closely related to fingering or flashing instability, which can initiate in a different way under a different test condition.

## DISCUSSION

### $\mu_{O_2}^{c}$ and $V^{c}$ for metal-insulator conversion

Charge neutrality, a large band gap, and disparate formation energies of different lattice defects rather strictly define the stoichiometry of insulating 8YSZ. [36,37] Therefore, pumping out more oxygen than letting into it unavoidably renders it unstable, decomposing into a suboxide starting at the cathode. Our study has established the suboxide is metallic, and the movement of the metal-insulator interface controls the resistance degradation, which ends in a total conversion into a metallic suboxide. (Because of fingering, the entire sample may not be metallic but at least there is a continuous stretch of metallic suboxide from cathode to anode.) Although the conversion speed is complicated by transport considerations in both phases and at both electrodes being sensitively dependent on their local geometries and atomistic kinetics, the conversion thermodynamics is straightforward and can be easily understood in terms of the phase rule using a simple model outlined in our theory. In this model, we estimated $PO_2^c$ and $\mu_{O_2}^c$ by the $PO_2$ and $\mu_{O_2}$ that raise the Fermi energy of electrons in the ionic YSZ to the Fermi level in the metallic phase. From the formation energy of $ZrO_2$, one obtains the electrolytic voltage of 2 V, which is the difference between the Fermi energy in Zr metal and that in $ZrO_2$. (This value much less than the band gap of YSZ, which at the standard state is p-type with a band gap of ~6 eV.) So the upper bound of $-\mu_{O_2}^c$ is 8 eV using Eq. (1a) that sets $1/2\Delta\mu_{O_2} + 2\Delta\mu_e = \Delta\mu_{O^{2-}} = 0$. The actual value may be evaluated by measuring the electromotive force (emf) of an open circuit, which is $(\mu_{O_2}'' - \mu_{O_2}^c)/4e$ after forming the metallic phase. However, at the lower temperatures used in our experiments, this emf is difficult to measure. But Janek and Korte reported an emf between 2.0 V and 2.5 V at 500°C in a sample with a completely blocked cathode [20], which is consistent with the above estimate.

While $\mu_{O_2}^c$ is a thermodynamic quantity, $V^c$ is completely kinetic in nature and ranges from zero with a completely blocked cathode to infinity with a zero-resistance cathode. The latter case corresponds to a $O^{2-}$ solid-electrolyte device with perfect electrodes: such device is transparent to the $O^{2-}$ flux. The former case of a completely blocked cathode will see immediate formation of the metallic phase at the cathode, because no matter how small the ionic current is, it will deplete oxygen and render some YSZ unstable. This was the case studied by Janek and Korte. [20] Although they did not report any quantitative data, they described that at 10 V a polycrystal YSZ had a black-front velocity about 1/10-th of that at 100 V, and 1/50-th of that at 500 V. These numbers are consistent with $V^c = 0$.

## Finger-like growth

Because $V^c = 0$, a small voltage such as 10 V was enough to drive black-front growth in Janek and Korte's experiment. [20] At this voltage, the growth velocity was very low and the growth front had a very slight curvature. However, at 100 V and 500 V, they also observed finger-like morphology from the very beginning of the experiments. This is consistent with our experience in which blackening invariably initiated with a finger or two.

At the tip of a relatively slender finger, the electric field, charge and $O^{2-}$ diffusion fluxes are all concentrated at the tip and have a radial dependence of *a*/(radial distance) where *a* is the tip radius. [29] This means that the effective transport length (*L* in our theory) is essentially reduced to *a*. Thus, the local electric field is much stronger and the tip growth rate much faster than predicted in our one-dimensional theory.

## Scaling laws for resistance degradation

Notwithstanding the above complications, our central prediction (**Fig. 3a**) that the resistance degradation curve scales with $t / R_o$ seems supported by the data (**Fig. 11**). This result is rederived below using the following simplified argument, which relates the growth of the metal-insulator interface to the $O^{2-}$ current in the insulating YSZ only. Equating the increment of metallic phase advancement $vdt = -Ldf$, which requires a charge removal rate of $2e\Delta C$ per unit volume when the volume is converted from YSZ to the metallic phase, to the charge transported by the $O^{2-}$ current density in YSZ during the same time increment, $\left| j^{\mathrm{YSZ}} \right| dt = (\sigma_{\mathrm{O}^{2-}} V / fL) dt$, we have $-(2e\Delta C) L df = (\sigma_{\mathrm{O}^{2-}} V / fL) dt$. Further approximating $f = r = R / R_{\mathrm{o}}$, we obtain $dr^2 / dt = -(V / R_{\mathrm{o}}) / (LAe\Delta C)$. Using values appropriate for 8YSZ and ZrO in our sample, we estimate at 325°C a characteristic degradation time of the order of $10^4$ s. We expect it shortened by about 10x because fingering enables a much shorter transport length than the sample length. Another 10x reduction is possible if $\Delta C$ is much smaller as will be discussed in the last subsection. These considerations will bring the expected degradation times into agreement with the observed ones.

The above result also captures the essence of **Fig. 3a** and **Fig. 11.** The stage I of the generic degradation curve should be a decreasing parabola, starting at its peak at $r = 1$ and $t = 0$, and ending in a steep drop to $r = 0$ at a $\tau$ of the order of $R_{\mathrm{o}} LAe\Delta C / V$. Because of this the $r - t/R_{\mathrm{o}}$ curves at different temperatures under the same $V$ and sample geometry ($L$ and $A$) should be invariant. Note, however, that $\tau$ will increase with $L$ even under the same nominal electric field $V/L$ and current density $V/AR_{\mathrm{o}}$, because when the same interface velocity $v$ obtains at the same field and current density, $dr/dt$ still diminishes with $L$ because $R_{\mathrm{o}} \sim L$. Therefore, at the same field and current density, the fractional resistance degradation is much more severe in thinner samples, and one must exercise caution when using the HALT data of bulk samples to predict the lifetime of thin devices, e.g., the forming time of resistance memory.

**Flash sintering**

The unfailing observation of a flash that ended the HALT is obviously relevant to flash sintering, which uses the flash to effect rapid densification of green bodies. In broader terms, flash sintering with the sample (a green body) placed under a constant voltage is practiced in two ways. In the first method, flash is brought about by a constant (furnace) heating rate, and we have shown that the sufficient condition for starting the flash [38,39], which we interpret as a thermal runaway [40], is when the Joule heating power exceeds the furnace heating power. (Materials studied in the literature all found their flash temperatures $T_{\text{on}}$ varying within a factor of two, so their furnace heating power being proportional to $T_{\text{on}}^4$ vary within a factor of 12.5 at $T_{\text{on}}$. Not surprisingly, their power densities at $T_{\text{on}}$ all lie within a narrow range as if there is a "universal" critical power density. [41]) A variation of this practice is to add auxiliary heating using a conducting die, which we also treated elsewhere. [42] In the second method [10], the sample is held at a constant temperature, and the flash occurs after a certain waiting time just as in our HALT. (Indeed, YSZ—though 3YSZ instead of 8YSZ—is the most popular material studied for flash sintering.) Our work makes clear that flash in the second method is also a thermal runaway phenomenon but it happens because the sample resistance has decreased by orders of magnitude by the YSZ-to-metal conversion, and what ignites the flash is the last "bridge" to the anode that bears nearly all the Joule heating power in the circuit. Note that if the flash is not immediately terminated or after the flash the sample remains at high temperature for an extended time, then all traces of metallicity including the black color will disappear during "zero-field cool," by reoxidation, as observed in our flashed samples. This speaks to the importance of the following designs in our study, *in situ* work and monitoring resistance change during cooling under various conditions, and the use of an intermediate temperature to minimize random diffusion of oxygen.

## Related work

Several related YSZ studies are briefly commented below. First, concerning flash sintering, Kirchheim proposed a theory, similar to that of Waser et al. for titanates [1-3], and suggested YSZ is a p-n junction with predominantly electronic conduction when both electrodes are ion-blocking allowing $O^{2-}/V_O^{\cdot\cdot}$ to redistribute. [43] Our oxygen bubbling experiment was expressly conducted to disprove it: anode is not blocking even at relatively low temperatures. Also, without incurring a phase transition it is thermodynamically impossible for YSZ to have more electrons/holes than $V_O^{\cdot\cdot}$ and become an electronic conductor. Grimley et al. interpreted Joule heating in a flashing 8YSZ in terms of electronic conductivity in the sample, which is again unrealistic without a phase transition for the same reason. [44] (This study focused on the time after the flash—which lasted 0.5 to 20 minutes, instead of the time to flash as we did. This is odd, for the appeal of flash sintering lies mainly in the possibility of a very short densification time, sometimes just seconds.) Third, West and coworkers [45,46] found a gradual (within a few minutes at 471-556°C) and large resistance decrease (2-5x) when 8YSZ was subject to a positive voltage (0-20 V) or an oxidizing atmosphere ($O_2$ vs. $N_2$), which they attributed to $O^{2-}$ oxidation near the anode. However, in **Fig. 9**, we did not observe any resistance decrease in the samples with a large $R^{\text{anode}}$, which would have made anode more oxidizing.

Our direct confirmation that oxygen depletion can easily convert YSZ to a metal is relevant to several melt processing methods involving high-energy heating and rapid quenching in an oxygen-lean environment, presumably losing much oxygen in the process. They include laser [23,47] and electron beam melting [48], yielding a bulk-like zirconia of a substantially similar crystal structure as 8YSZ, but with a distinct black color, severe narrowing of optical band gap, and metallic DC conductivity that also reverts back to insulating upon modest reheating in air. [23] First-principles calculation motivated by the observations found a much smaller band gap in $ZrO_{1.75}$ than in $ZrO_2$. [23] In our experiments, along the length between cathode and anode more than one suboxide may well exist, in the order of increasing oxygen content and separated by a set of two-phase interfaces following the phase rule. Importantly, because their oxygen contents are closer to that of 8YSZ than the monoxide we assumed, their $\Delta C$ is smaller and less oxygen depletion is needed for their growth, hence a faster degradation rate is expected. Also important to note is the role of rapid quenching, without which the suboxides would have reoxidized during cooling from the melt state. In this regard, it seems plausible that during forming of resistance memory, rapid quenching made possible by the nanoscale sample size and the low forming temperature can more readily retain the metallic phase. This again speaks to the need of using relatively low temperatures in our experiment. Indeed, at higher temperatures, because of ready exchange with atmospheric oxygen, similar electrical tests failed to produce any resistance degradation even when conducted at much larger ionic current densities and with evidence of strong electrode polarization. [31,32,49,50]

## CONCLUSIONS

(1) A thermodynamic and kinetic theory for resistance degradation under a DC voltage is proposed for 8YSZ, which does not have totally ion-blocking electrodes as in perovskite titanates.

(2) Above a kinetics-dependent critical voltage, 8YSZ loses oxygen and turns metallic starting from the cathode side. At larger voltages, the metallic phase takes the form of fingers that grow and coarsen, turning YSZ black and eventually providing a completely metallic path ending the test in a thermally runaway flash.

(3) Besides temperature and electric field, the largest effect on degradation kinetics comes from anode kinetics and initial resistance. Size is important: at the same temperature and electrical field, thin films degrade proportionally faster.

(4) Part or all of the metallic phase may undergo a diffusionless metal-insulator transition below a critical temperature $T_c$, which is state dependent, reflecting the precarious nature of the metallic phase in a highly defective YSZ. In a field-dependent manner, black YSZ can also revert to a colorless insulator by defect/electron/hole diffusion at a modest temperature. Therefore, post mortem examination may not reveal the metallic character of the post-degradation state.

(5) While there is a large scatter in degradation kinetics, the degradation time generally scales with the initial resistance, and the scatter becomes smaller when repeating the same test in the same sample. However, restarting the test in the same sample under a different test condition will reintroduce the large scatter, which strongly suggests that degradation kinetics are inherently statistical because of fingering and flashing instability.

## Figures

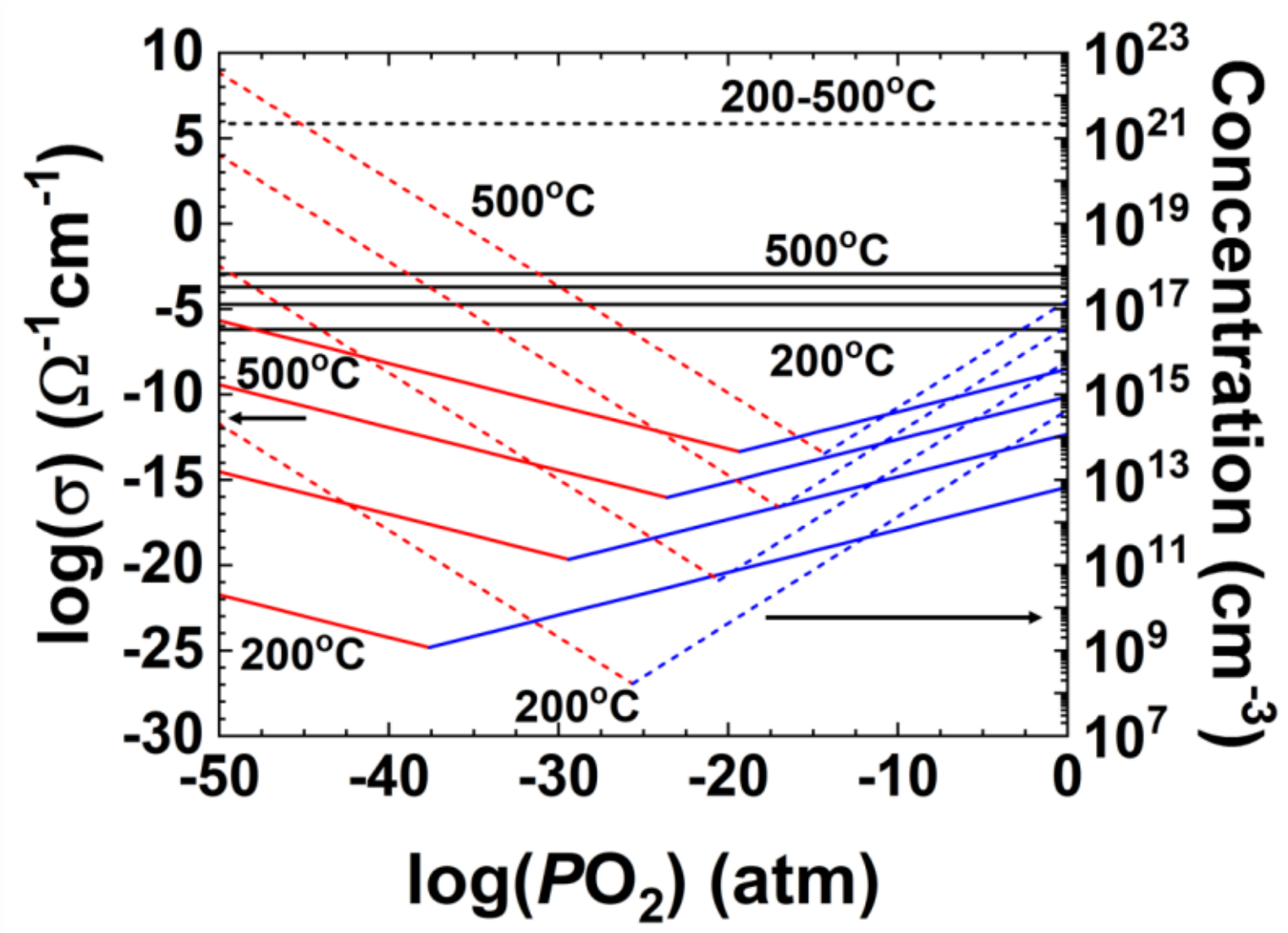

**Fig. 1** Log ionic (black), electron (red) and hole (blue) conductivity of 8YSZ versus log $PO_2$ at four temperatures. Corresponding concentrations of oxygen vacancies, electrons and holes shown in broken lines. Modified from Park & Blumenthal. [14]

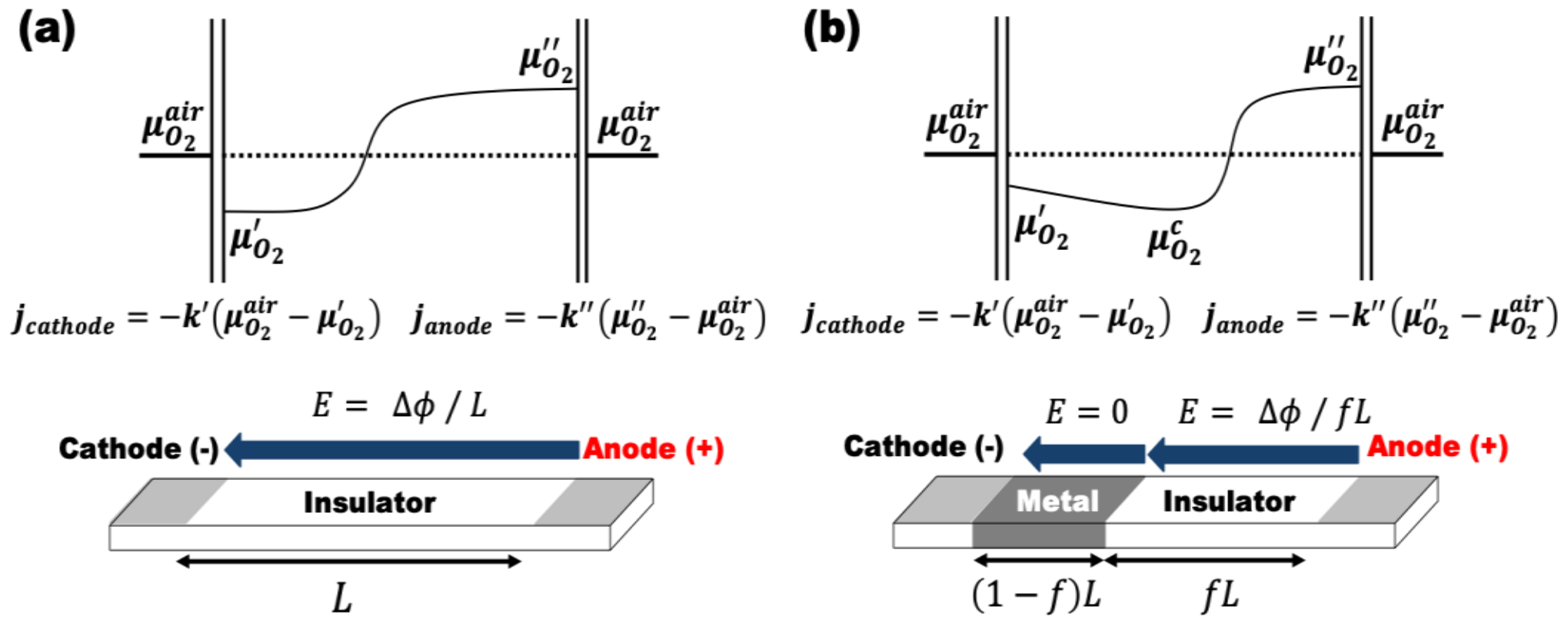

**Fig. 2** Schematic phase and electrode configuration, electric field, electrode kinetics and oxygen potential distribution across the sample when (a) $V < V^c$ and (b) $V > V^c$.

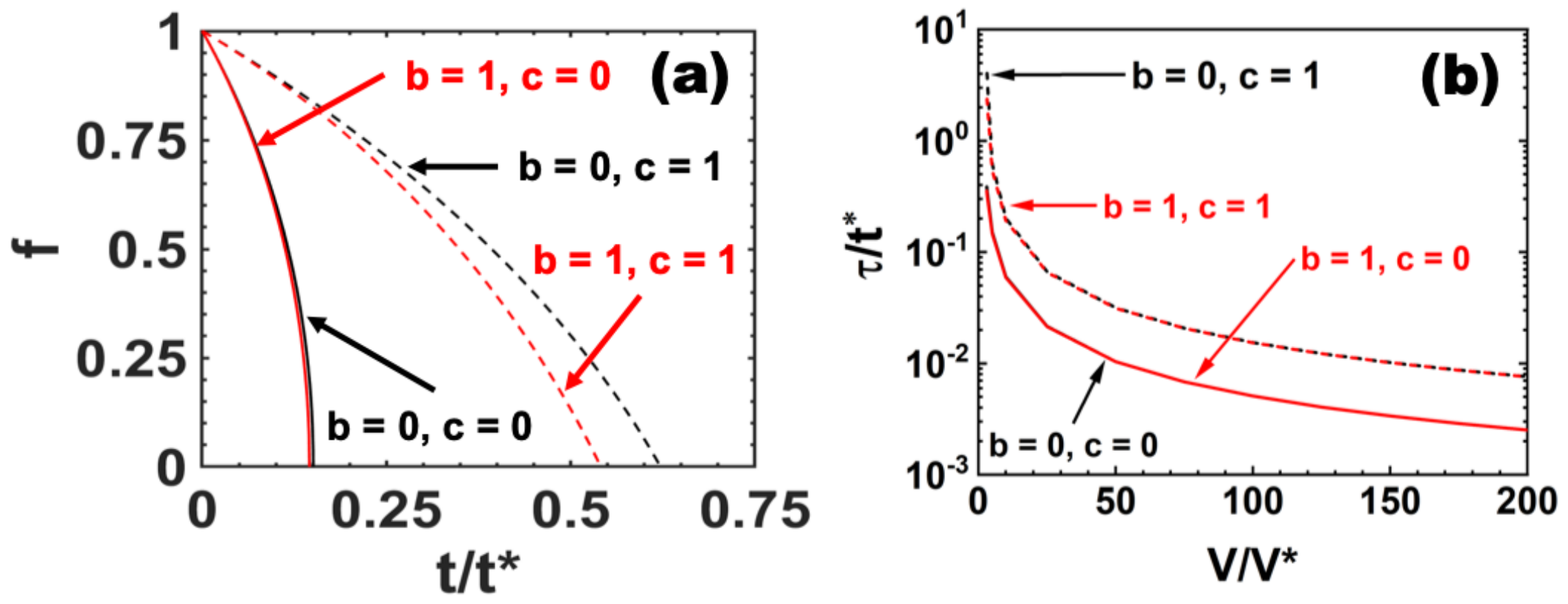

**Fig. 3** Predicted (a) $f$ vs. reduced time and (b) $\tau$ vs. reduced voltage with parameter *c* (i.e., $R^{\text{anode}}/R^{\text{YSZ0}}$) having a much stronger effect than parameter *b*. Note: $f$ is essentially the fractional (or normalized) resistance, $R/R_o$. In all cases $a = 1$.

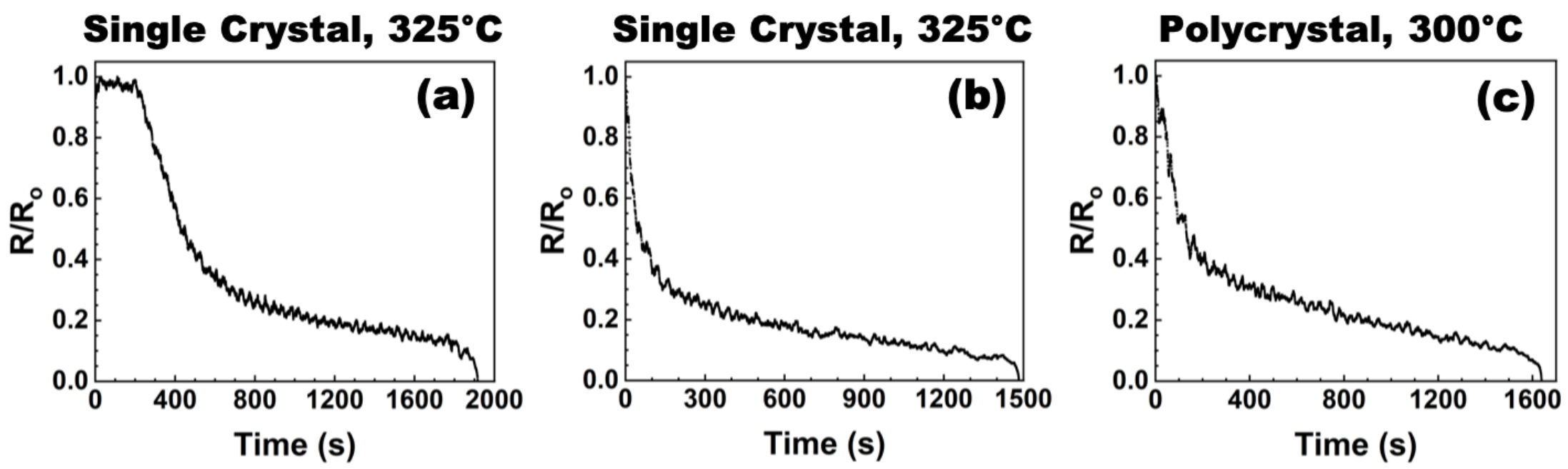

**Fig. 4** Representative degradation curves in $R/R_o$ and time. Stage I is most evident in (a), least in (b). All samples tested under 200 V. (a) has Pt wire as anode. (b) and (c) use Pt paste on both electrodes.

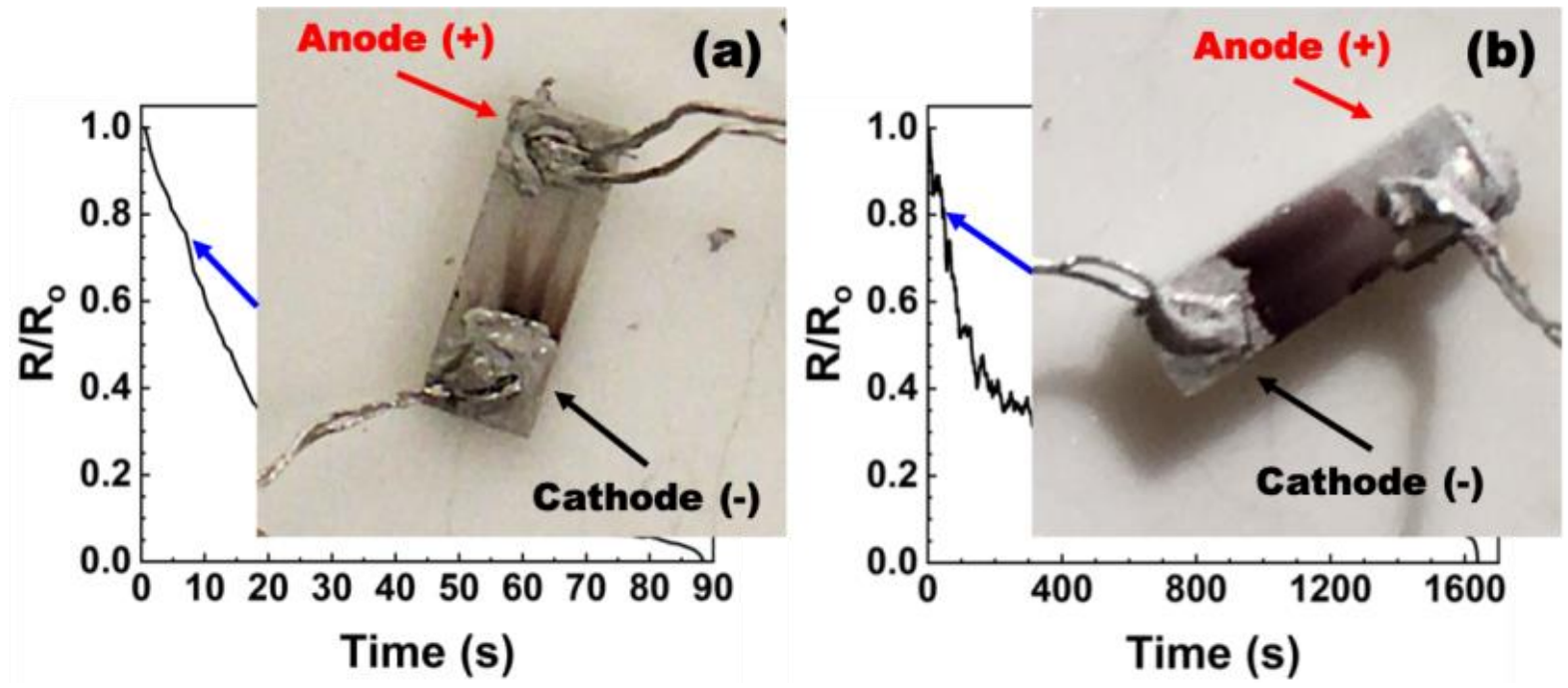

**Fig. 5** Blackening photographed at early stage of degradation in (a) a single crystal tested at 335ºC and 200 V to $r = 0.75$ with dark fingers already reaching the anode, and (b) a polycrystalline tested at 300ºC and 200 V to $r = 0.8$ with almost complete blackening.

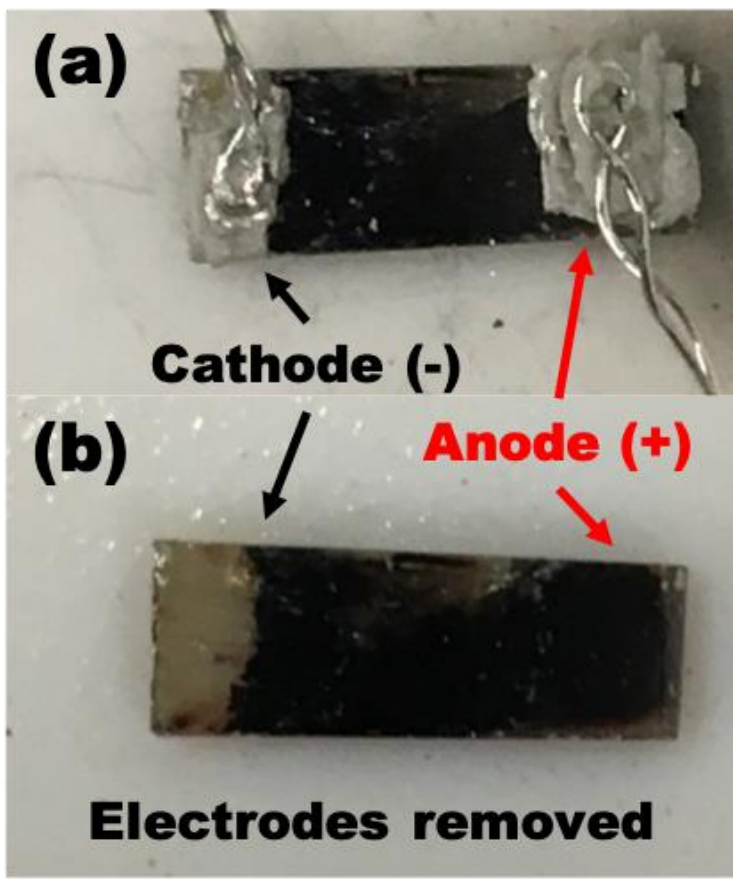

**Fig. 6** Photographs of a single crystal degraded at 350°C and 200 V, then field cooled under 200 V to room temperature, shown with Pt electrodes (a) intact and (b) removed at room temperature. A dark region exists beneath the anode but not the cathode.

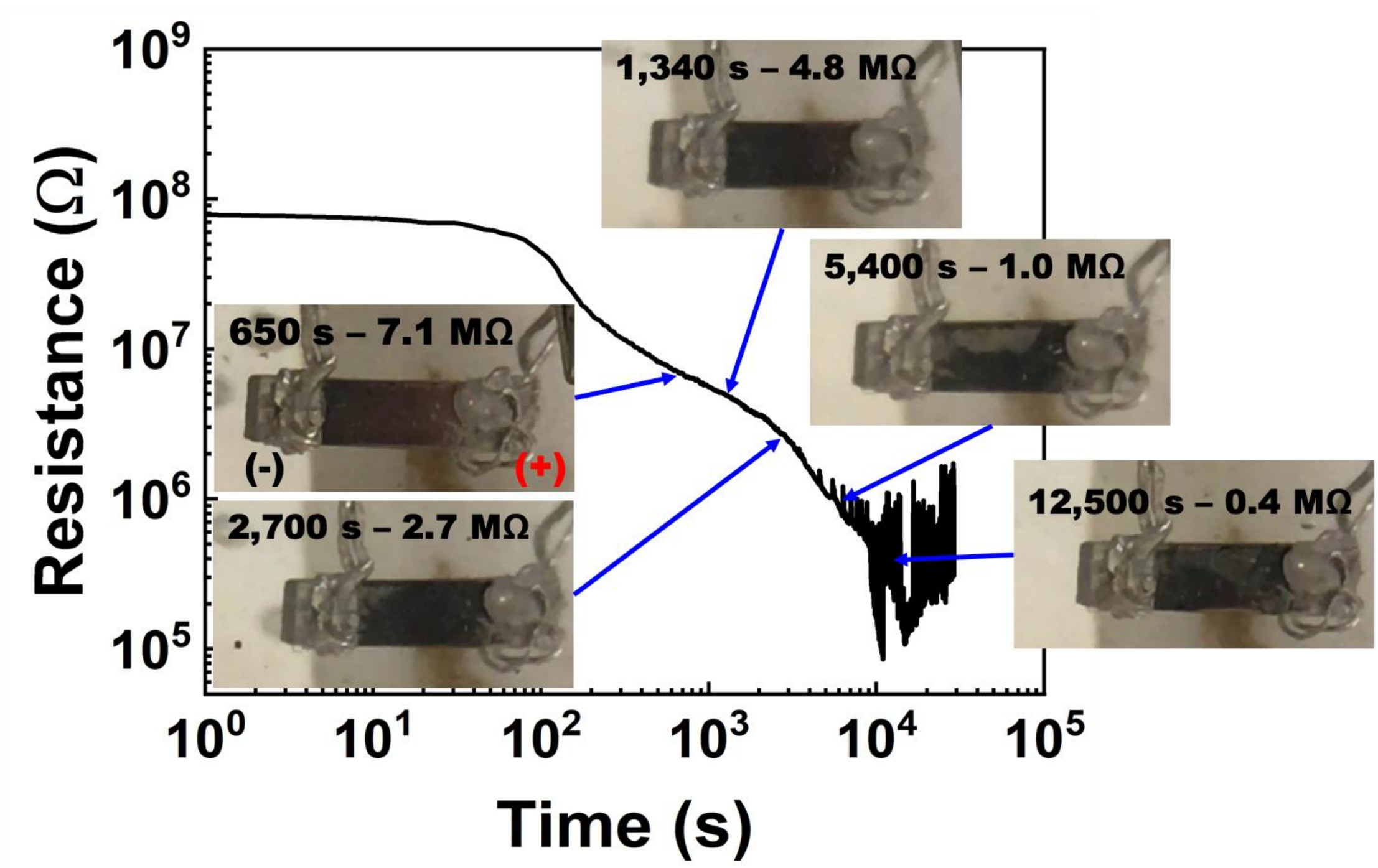

**Fig. 7** Serial photographs of a single crystal taken during a degradation test conducted in silicone fluid at 280ºC under 200 V. Already completely black at 1340 s, the crystal developed a white band near the upper edge of cathode by 2700 s that continued to grow in size at later time (5400 and 12500 s). Increased "noise" in electrical signal is due to convection disturbances developing over time.

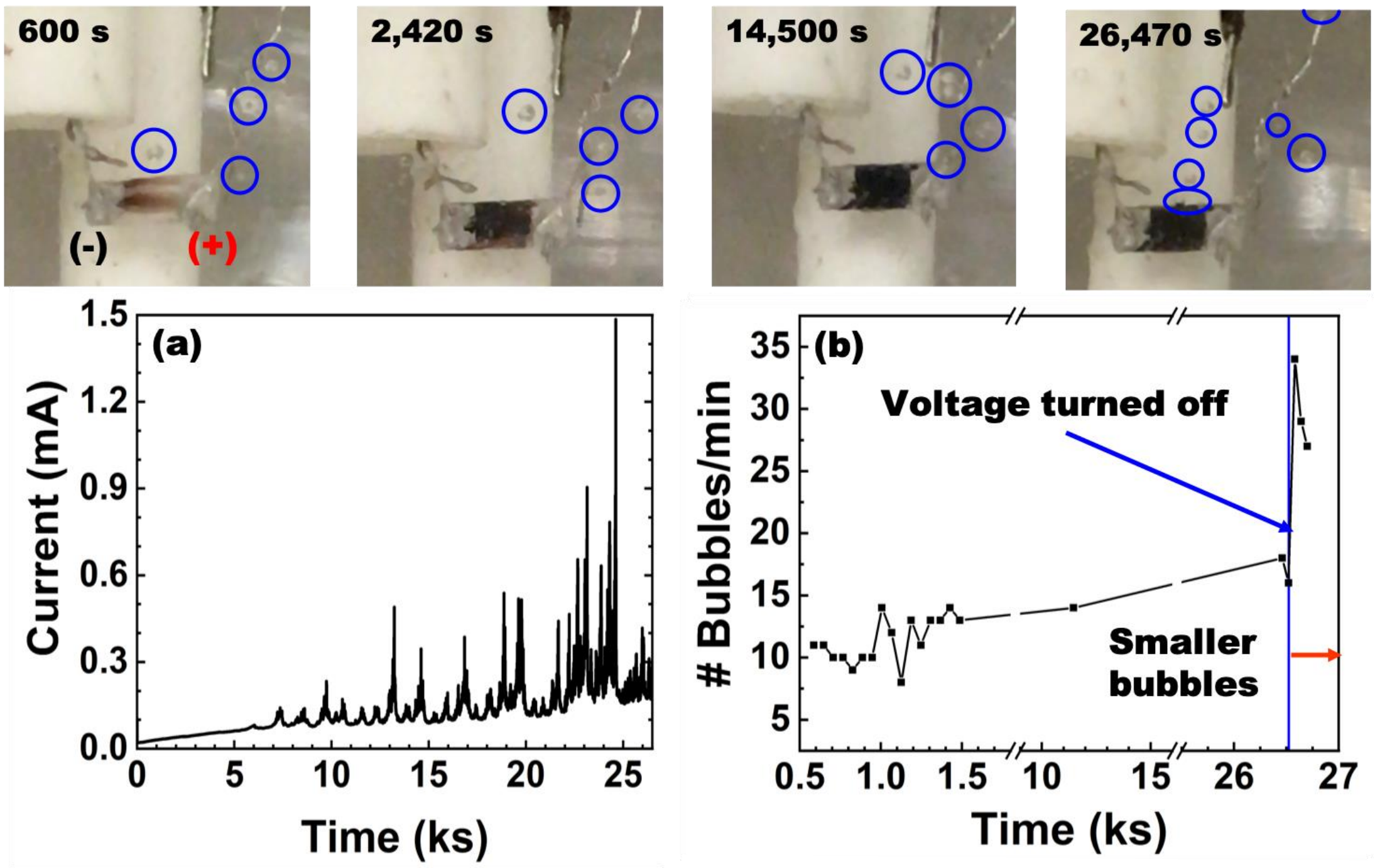

**Fig. 8** Serial photographs of a single crystal taken during a degradation test (see current-time trace on lower left) conducted in silicone fluid at 260°C under 200 V. Bubbles (circled) found in all photographs, increasing in numbers with time (see number statistics on lower right) but some even appearing after electric field was off. Although bubbles all originated from the anode, they often were attached to the sample edge, growing and migrating until becoming big enough to detach. Increased "noise" in electrical signal is due to convection disturbances developing over time.

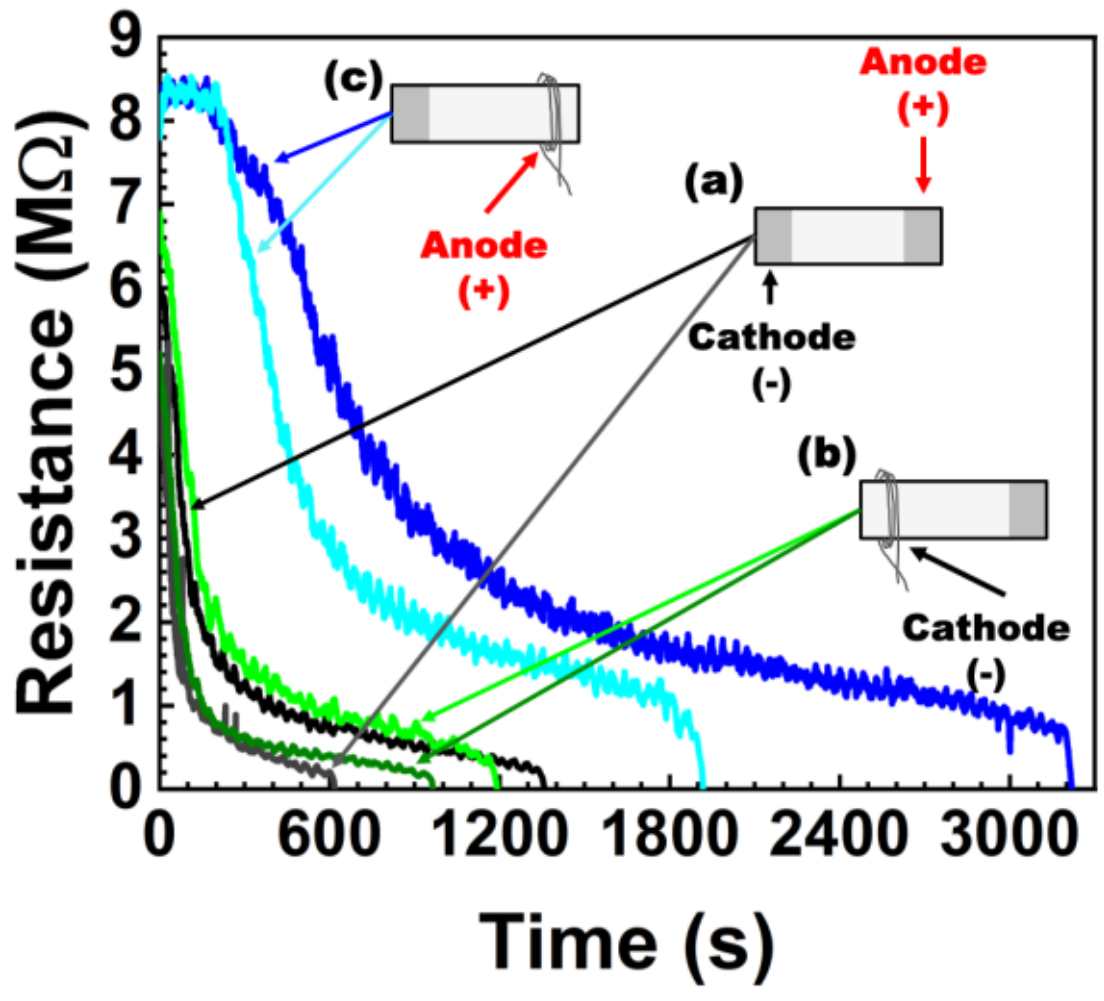

**Fig. 9** Resistance vs time plots for 6 single crystals samples tested at 325°C and 200 V with 3 different electrode configurations schematically shown by insets.

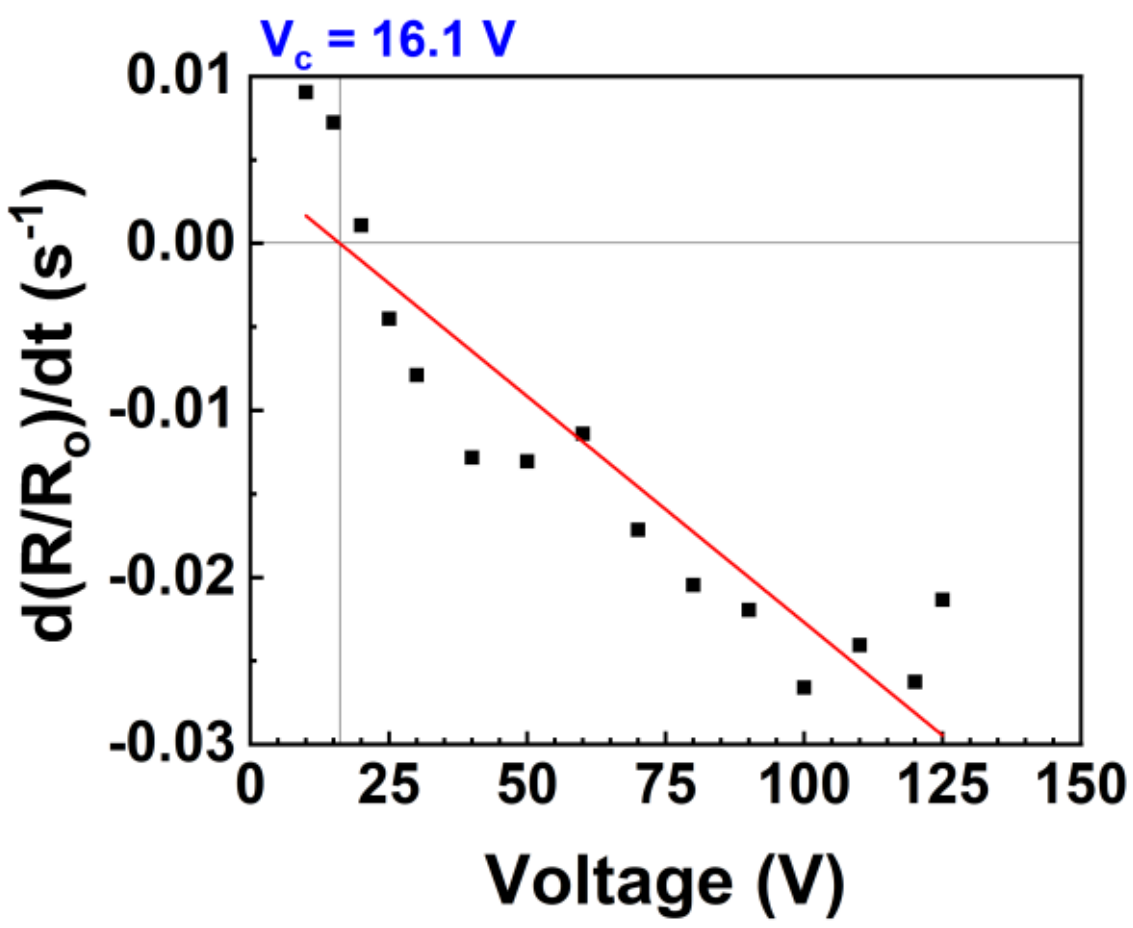

**Fig. 10** Average degradation rate $dr/dt$ during the first 20 seconds of degradation of a single crystal at 325°C and 15 to 125 V. The range of $r$ covered was from 1 to 0.33, with most from 1 to 0.5. No degradation apparent at $V < V^c = 16.1\,\mathrm{V}$.

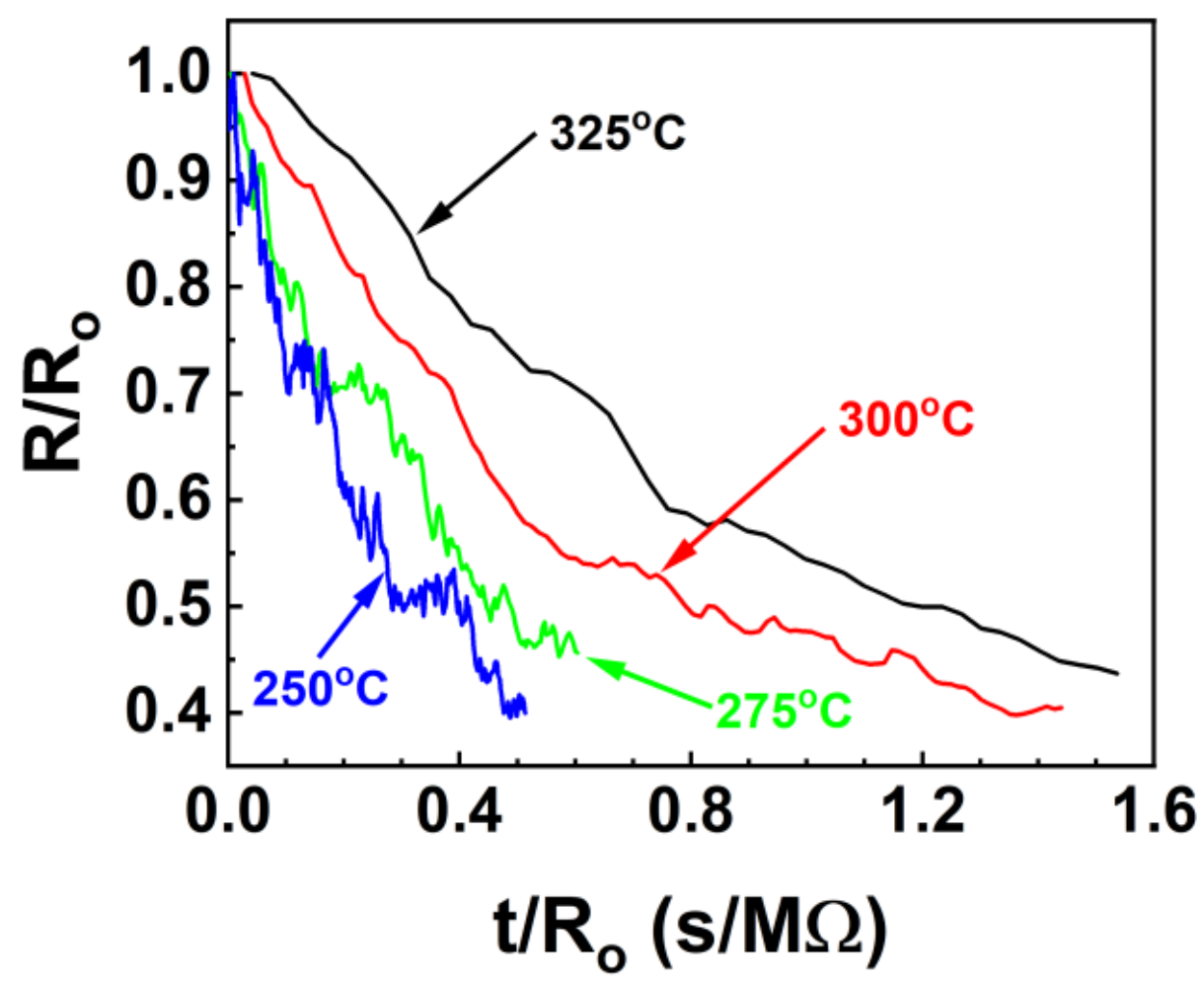

**Fig. 11** Normalized resistance $r$ versus normalized time, $t/R_o$, curves for a single crystal tested at 125 V at 4 temperatures.

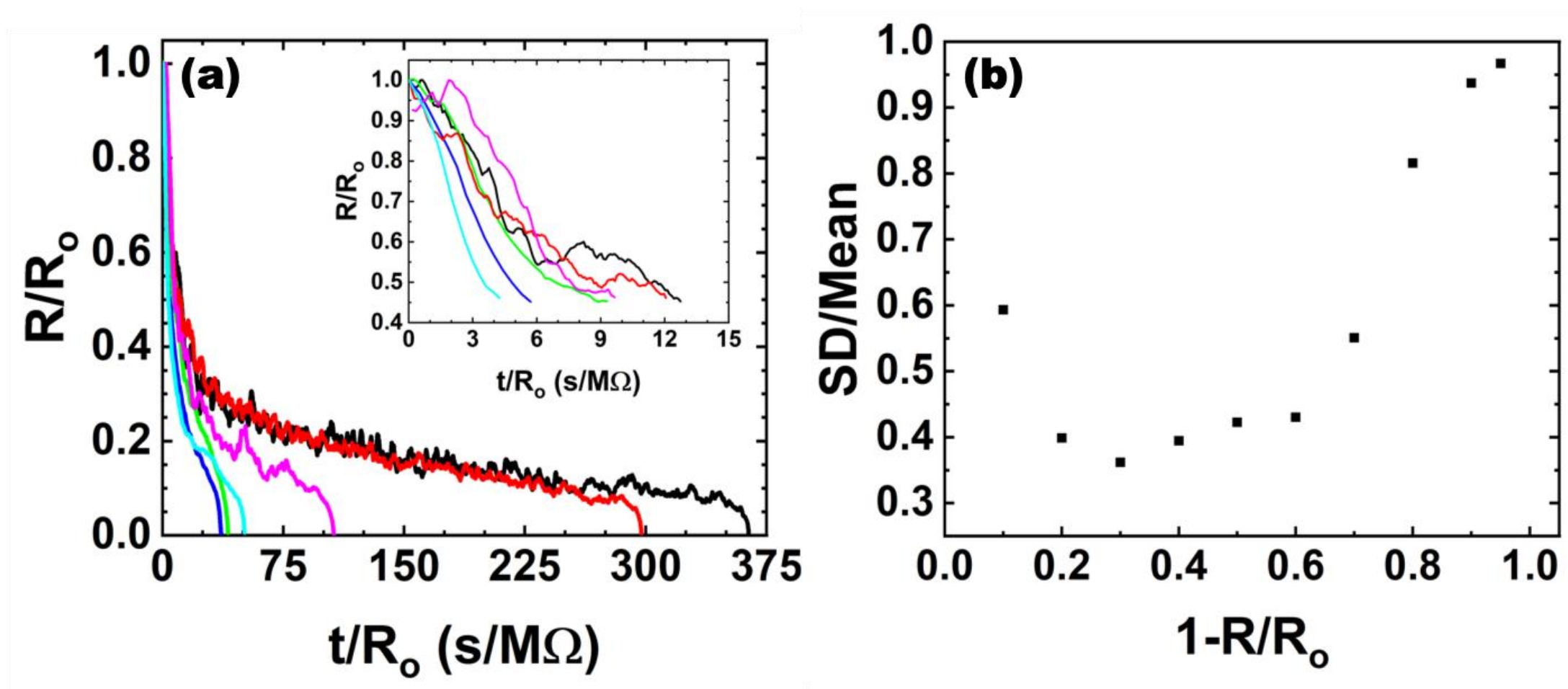

**Fig. 12** (a) Normalized resistance $r$ versus normalized time $t/R_o$ for 6 samples (single crystal) tested at 325ºC and 200 V. Inset: same curves from $r = 1$ to 0.45. (b) Ratio of standard deviation (SD) to mean of the normalized time at a given $r$.

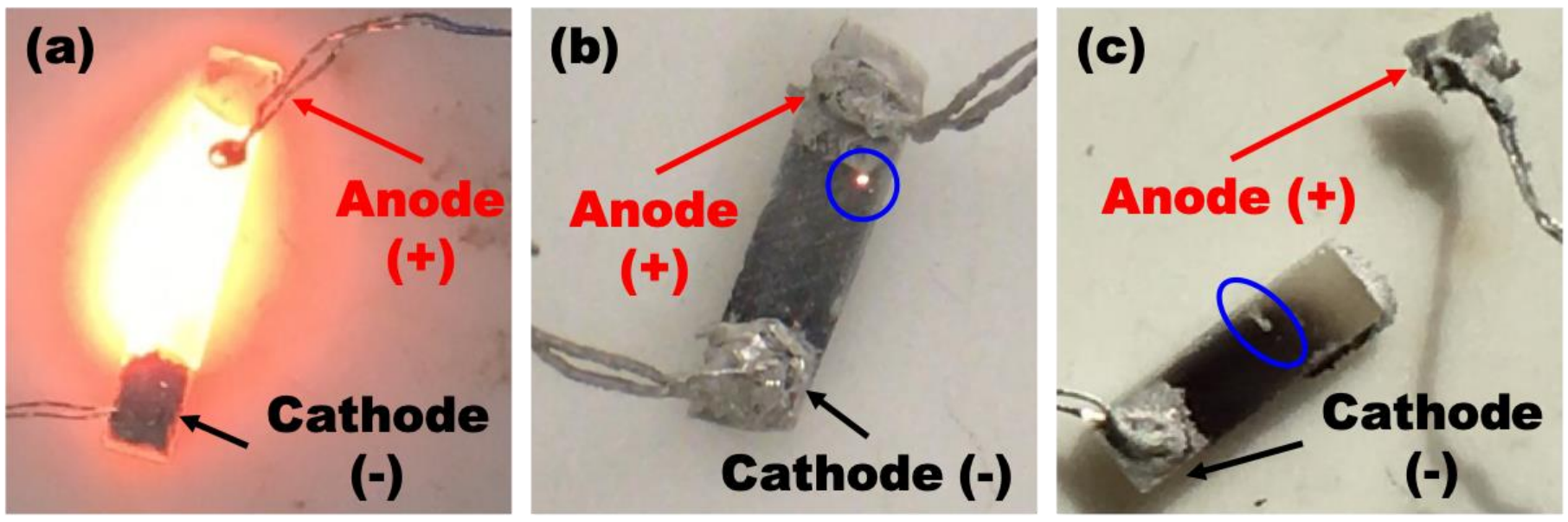

**Fig. 13** (a) A single crystal tested at 425°C lighting a flash throughout the entire area between the electrodes. This sample was transparent after cooling. (b) A polycrystalline sample tested at 335°C forming a hot spot next to a protrusion on the anode. (c) A polycrystalline tested at 300°C leaving a white spot in an otherwise black sample quenched after initial flash, at the white-spot location, because of anode detachment.

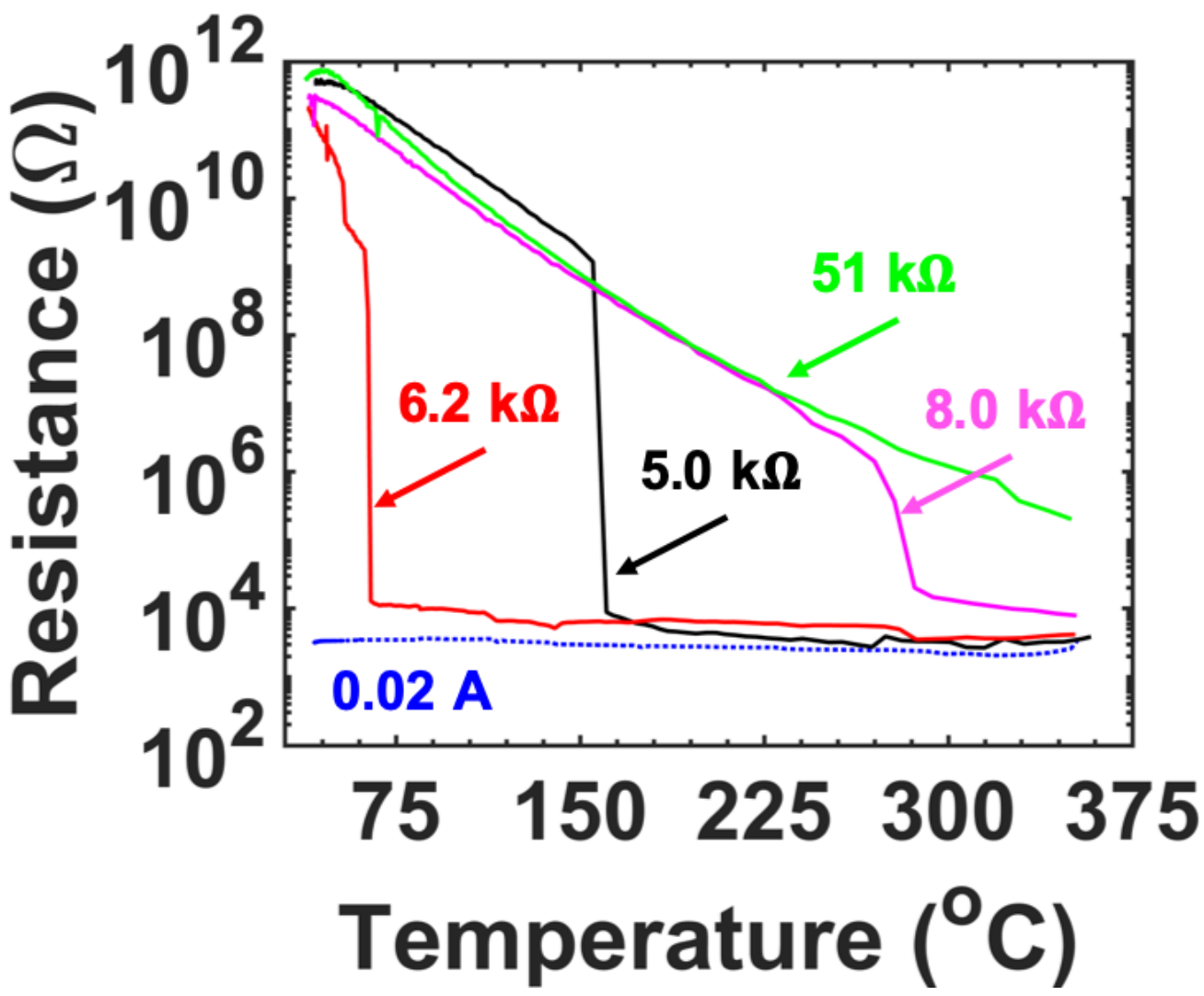

**Fig. 14** Temperature dependence of the resistance of a degraded (200 V at 350°C) single crystal cooled after resistance reaching $2R_L$, hence with a flash at $R_L < 7\ \text{k}\Omega$ or without a flash at $R_L > 7\ \text{k}\Omega$. Also shown is the dependence when cooled after current reaching a compliance limit of 0.02 A, which was maintained during cooling.

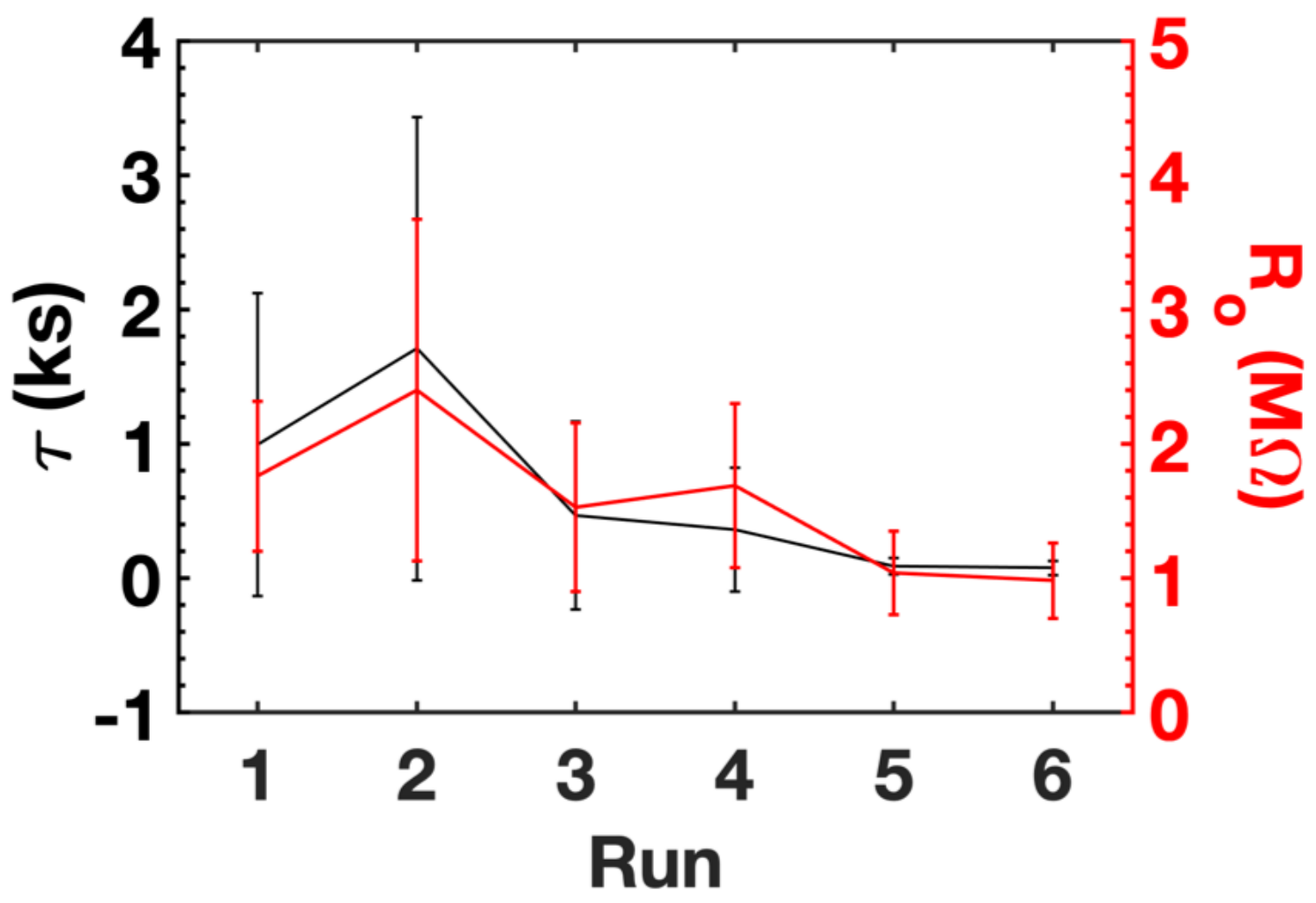

**Fig. 15** Average and scatter (bar) of $R_o$ (red) and $\tau$ (black) to reach the compliance limit of 0.02 A for 15 8YSZ single crystals repeatedly tested at 325°C and 200 V for 6 times. Scatter in both $\tau$ and $R_o$ decreases with test run.

# Supplementary information

## DC Electrical Degradation of YSZ: Voltage Controlled Electrical Metallization of A Fast Ion Conducting Insulator

Ana Alvarez[1], Yanhao Dong[2], I-Wei Chen[1]

[1] *Department of Material Science and Engineering, University of Pennsylvania, Philadelphia, PA 19104, USA*

[2]*Department of Nuclear Science and Engineering, Massachusetts Institute of Technology, Cambridge, MA 02139, USA*

**Table of Content**

**Theory of insulator-metal phase transition under a constant voltage**

**Phase equilibrium between zirconia and metallic suboxide**

We first ignore Y since cations are essentially immobile at low temperatures, so they have identical composition and distribution in different phases. Without loss of generality we consider only one metallic suboxide, the monoxide, $ZrO_{1+\delta}$ with some $O^{2-}$ interstitials, and one insulating oxide $ZrO_{2-\varepsilon}$, with some $O^{2-}$ vacancies. Under an $O_2$ gas of a partial pressure $PO_2$, the three-phase $(P=3)$ equilibrium obeys the phase rule, $P=C-F+2$, with two components (Zr and O) $C=2$ thus allowing only one degree of freedom *F*, which is the $PO_2$ isotherm, namely the *P-T* phase-coexistence line. Therefore, at each temperature, there is a unique $PO_2^c$, or equivalently a unique chemical potential $\mu_{O_2}^c$, at the three-phase equilibrium.

We may derive $PO_2^c$ and $\mu_{O_2}^c$ using a simple model. First, we consider the oxidation reaction

$$\frac{1}{2}O_2 + 2e = O^{2-} \qquad \text{(A1)}$$

which dictates the equilibrium condition for the chemical potential, $\mu_i$ of the *i*-th species,

$$1/2\,\mu_{O_2} + 2\mu_e = \mu_{O^{2-}} \qquad \text{(A1a)}$$

and the electrochemical potential $\tilde{\mu}_i$,

$$1/2\,\mu_{O_2} + 2\tilde{\mu}_e = \tilde{\mu}_{O^{2-}} \qquad \text{(A1b)}$$

In the above, $\tilde{\mu}_i \equiv \mu_i + z_i e\phi$, where $\phi$ is the electric potential, $z_i e$ the charge of the *i*-th species, and $e$ the elementary charge. Second, we note that the population of delocalized electrons is very large in the stoichiometric ZrO, which is metallic. So its Fermi level, which is the chemical potential of electron, $\mu_e$, is not affected by the above reaction. Therefore, any $O^{2-}$ interstitials inserted into the suboxide must follow $\mu_{O^{2-}} = 1/2\,\mu_{O_2} + \text{constant}$ according to Eq. (A1). Third, we now include $Y^{3+}$ as an acceptor dopant into $ZrO_2$ forming YSZ with a large dopant concentration that fixes the concentration of oxygen vacancies to maintain charge neutrality. So, any $O^{2-}$ ions inserted into YSZ filling an oxygen vacancy can hardly affect $O^{2-}$ concentration, i.e. $\mu_{O^{2-}} = \text{constant}$. Fourth, the chemical potential of electron in YSZ follows $\mu_e = -1/4\,\mu_{O_2} + \text{constant}$ to Eq. (A1a), so its Fermi level increases with decreasing $\mu_{O_2}$. Fifth, at a sufficiently low $PO_2^c$ or $\mu_{O_2}^c$, the Fermi level of electrons in YSZ will equal the constant Fermi level of the metallic suboxide, which determines $PO_2^c$ and $\mu_{O_2}^c$. Note that although the phase rule dictates that two-phase equilibrium must occur at a unique $PO_2^c$ and $\mu_{O_2}^c$, inside the single-phase, metallic suboxide, the $\mu_{O_2}$ can actually be higher, meaning a higher concentration of $O^{2-}$ interstitials; this turns out to be the case at the metal-cathode interface as will become clear later.

In summary, in the metallic phase, $\mu_{O^{2-}}$ varies with $\mu_{O_2}$ but $\mu_e$ is fixed, and vice versa in the ionic phase. The two phases share a unique $\mu_{O_2}^c$ at their interface, and once the $\mu_{O_2}$ distribution on two sides of the interface is known, the distributions of $\mu_{O^{2-}}$ and $\mu_e$ are also known from Eq. (A1b). These are all the chemical potentials needed to solve the transport and phase-equilibrium problem—no need for the chemical potentials of cations because they are immobile at the temperatures of typical HALT and device operations.

**Critical voltage for the formation of metallic phase**

Consider a one-dimensional slab of YSZ placed in air. Across its length $L$ there is a voltage $V$ driving a (negative) current flowing from the right (anode) to the left (cathode). Below we will determine the critical voltage $V^{c}$ below which no metallic suboxide forms. The metallic phase grows at above $V^{c}$ from the cathode and it eventually reaches the anode. The boundary conditions are those at the two electrodes: $O^{2-}$ enters from the cathode and exits from the anode; no oxygen allowed to enter or exit from the side.

Given the fixed composition of YSZ, the only way to start forming metallic suboxide is to lose more oxygen at the anode than gain oxygen at the cathode. This is clear from mass balance. Below we will solve the $\mu_{O_2}$ distribution that satisfies the thermodynamics and kinetics in support of the above picture. Our experiment strongly suggests that the anode can quite freely release $O_2$, which may not be the case for the cathode. So we model these kinetics at these interfaces as interface-controlled ones of the simplest kind having a current density proportional to the driving force, which is the difference of $\mu_{O_2}$ across the interface. At the cathode, we set the limit of the lowest $\mu_{O_2}$, called $\mu'_{O_2}$, at $\mu^{c}_{O_2}$ when the metallic phase forms, but set no limit for the anode $\mu_{O_2}$, called $\mu''_{O_2}$. This gives the following current densities, which are negative,

$$j^{\text{cathode}} \equiv j' = -\text{Min}\left\{k'(\mu^{\text{air}}_{O_2} - \mu'_{O_2}), k'(\mu^{\text{air}}_{O_2} - \mu^{c}_{O_2}) \equiv \left|j'_{\max}\right|\right\} \quad \text{(A2a)}$$

$$j^{\text{anode}} \equiv j'' = -k''(\mu''_{O_2} - \mu^{\text{air}}_{O_2}) \quad \text{(A2b)}$$

In the above, $k'$ and $k''$ are rate constants and $j'_{\max}$ is set by the thermodynamic limit.

Next, we consider the current density in YSZ. Since $\mu_{O^{2-}}$ is constant, the force to drive an oxygen current come from the electrical field *E* only, which must be constant at the steady state. So it can be obtained from the electric potential difference $\Delta\phi$ between the anode and the cathode to give the steady-state current density

$$j^{YSZ} = -\sigma_{O^{2-}}\Delta\phi / fL \quad \text{(A3)}$$

Here, $f$ is the fraction of the length that is YSZ; for now, $f = 1$. Since it is actually the voltage *V* that controls the experiment, we need to relate $\phi$ to *V*, which is the difference between the Fermi level—namely the electrochemical potential of electron—of the cathode and the anode. From Eq. (A1b), this difference is

$$\Delta\tilde{\mu}_e = -\Delta\phi - (\mu''_{O_2} - \mu'_{O_2})/4e \quad \text{(A4a)}$$

In writing the above, we recognized that $\mu_{O^{2-}}$ is constant. So,

$$V = -\Delta\tilde{\mu}_e = \Delta\phi + (\mu''_{O_2} - \mu'_{O_2})/4e \quad \text{(A4b)}$$

Finally, we use the continuity equation to solve the problem. As shown in **Fig. 1**, before reaching $|j'| < j'_{max}$ at the cathode, the continuity equation is

$$-k'(\mu^{air}_{O_2} - \mu'_{O_2}) = -\sigma_{O^{2-}}\Delta\phi / L = -k''(\mu''_{O_2} - \mu^{air}_{O_2}) \quad \text{(A5)}$$

The solution of Eq. (A4b) and Eq. (A5) for the following oxygen potentials is

$$-k'(\mu'_{O_2} - \mu^{air}_{O_2}) = k''(\mu''_{O_2} - \mu^{air}_{O_2}) = (1 + \sigma_{O^{2-}}/4eLk' + \sigma_{O^{2-}}/4eLk'')(\sigma_{O^{2-}}V/L) \quad \text{(A6)}$$

We can also express *V* in terms of the current density using $j = j' = j'' = j^{YSZ}$

$$V = (|j|L/\sigma_{O^{2-}})(1 + \sigma_{O^{2-}}/4eLk' + \sigma_{O^{2-}}/4eLk'') \quad \text{(A7)}$$

Here, the first bracket on the right is just the Ohm's law, and the second bracket further includes the equivalent cathode and anode resistance, $1/4eLk'A$ for $R^{cathode}$ and $1/4eLk''A$ for $R^{anode}$, divided by the resistance of YSZ of a unit area, $L/\sigma_{O^{2-}}$. Such form is well known in the transport problems where interface kinetics is limited. In the simple case of constant $k'$ and $k''$, it only adds a term to the total resistance, which is still Ohmic, meaning that the resistance is independent of $V$. This breaks down when $k'$ and $k''$ themselves depend on $\mu'_{O_2} - \mu^{air}_{O_2}$ and $\mu''_{O_2} - \mu^{air}_{O_2}$, as in the so-called Butler-Volmer kinetics. Nevertheless, as long as the electrode resistance is small compared to that of YSZ, e.g., when $L$ is long and $k'$ and $k''$ are large, the device resistance is still dominated by the YSZ resistance and appears relatively Ohmic.

The critical voltage $V^c$ is set by replacing $j$ by $j'_{max}$,

$$V^c = (k'(\mu^{air}_{O_2} - \mu^c_{O_2})L/\sigma_{O^{2-}})(1 + \sigma_{O^{2-}}/4eLk' + \sigma_{O^{2-}}/4eLk'') \qquad (A8)$$

which increases with $k'$ because of a larger $j'_{max}$ but decreases with $k''$ because of a lesser anode resistance. So the critical voltage is very sensitive to the electrode kinetics and electrode conditions (roughness, porosity, microstructure, composition, etc.) and best left as an empirical property to be determined experimentally.

When the $V$ exceeds $V^c$, the $\mu'_{O_2}$ may increase or decrease from its value $\mu^c_{O_2}$ at $V = V^c$. First consider $\mu'_{O_2} < \mu^c_{O_2}$, which means a larger driving force at the cathode but the current density is still set at $j'_{max}$. This also means the apparent resistance increases, which we have not seen in our experiment, nor are we aware of such data in the literature. (The resistance occasionally does rise in the beginning of a constant voltage experiment, but only by a small amount and for a short time, before it gradually levels off and starts to drop. This is not what is expected from the above scenario.) The other possibility is $\mu'_{O_2} \geq \mu^c_{O_2}$, which means a smaller driving force and a smaller current density, which allows the possibility of resistance decrease as indeed seen in our experiment. This scenario will be considered in more detail in the next section.

In the above, we did not specify the electronic current of electrons and holes. In YSZ, the electrolytic range, which is the range of $\mu_{O_2}$ over which the electronic conductivity $\sigma_{eh} \equiv \sigma_e + \sigma_h$ is small compared to $\sigma_{O^{2-}}$, is very wide, especially at low temperatures. Therefore, there is little error in ignoring the electronic current in the above solution. However, as we emphasized elsewhere, the $\mu_{O_2}$ distribution in YSZ is sensitively tied to $\sigma_{eh}$, which has a minimum at an intermediate $\mu_{O_2}$ where the potential experiences a step-like change. This minimum conductivity essentially coincides with the intrinsic electronic conductivity of an undoped $ZrO_2$, which is very small because of a very wide bandgap. It can be shown that the step-like rise in $\mu_{O_2}$ is needed to maintain a steady-state electronic current, however small it is, across the entire YSZ. This happens at the kinetic bottleneck where the conductivity minimum resides.

**Growth of the metallic phase and resistance degradation**

The situation of $V > V^c$ is depicted in **Fig. 2**, which shows a metallic phase extending from the cathode over a length of $(1-f)L$, and YSZ on the right over a length of $fL$. As time $t$ increases, $f$ decreases, and the phase boundary moves to the right at a time-dependent (or position/*f*-dependent) velocity *v*. Because the resistance variation is mostly that of the YSZ phase, it is proportional to $df/dt$. So, the plot of $f(t)$ is a good indicator of the resistance degradation. To solve *v* and $f$, we again consider current continuity, which now includes the metallic phase. It will become obvious that with time the growth and the resistance degradation will accelerate.

Since there is no electric field in a metal, its (interstitial) $O^{2-}$ can only move by diffusion. With the distance from the anode denoted by $x$, we use the Fick's first law for the $O^{2-}$ flux, $-c(D/k_BT)\partial\mu_{O^{2-}}/\partial x$ and $\mu_{O^{2-}} = 1/2\,\mu_{O_2} + \text{constant}$ in the suboxide to obtain the $O^{2-}$ current density, $j^{\text{metal}} = ec(D/k_BT)\partial\mu_{O_2}/\partial x$. Here, $D$ is the self-diffusivity of $O^{2-}$ interstitial, $c$ is its concentration, and $k_BT$ has its usual meaning. Since the suboxide is likely to be defect-rich and a fast oxygen conductor itself, we will assume that $c$ is a constant. Therefore, at the steady state $\partial\mu_{O_2}/\partial x$ must be a constant and may be replaced by $\Delta\mu_{O_2}/(1-f)L = (\mu_{O_2}^{c} - \mu'_{O_2})/(1-f)L$. So, the current density in the metallic phase is

$$j^{\text{metal}} = ec(D/k_BT)(\mu_{O_2}^{c} - \mu'_{O_2})/(1-f)L \equiv k^{\text{metal}}(\mu_{O_2}^{c} - \mu'_{O_2})/(1-f)L \qquad \text{(A9)}$$

The above diffusion transporting $O^{2-}$ in the metallic phase, from left (the cathode side) to right (the YSZ side), demands that the $\mu_{O_2}$ is higher at the cathode and lower at the metal-YSZ interface, which is consistent with our expectation mentioned in the last section. Indeed, as the metallic phase grows with time, $\mu'_{O_2}$ must increase further to counter a longer diffusion distance. So, counterintuitively, the cathode is not the most reducing location in the slab, and it actually "reoxidizes" over time after the metallic suboxide forms. This prediction did find support in our experiment. It also implies that the magnitude of the cathode current density reaches the maximum $|j'_{\max}|$ at $V^{c}$ and decreases at $V > V^{c}$.

The continuity of current density is maintained across the electrode-metal and the YSZ-anode phase interface, which gives

$$-k'(\mu_{O_2}^{air} - \mu'_{O_2}) = k^{metal}(\mu_{O_2}^{c} - \mu'_{O_2})\big/(1-f)L = j' \qquad \text{(A10a)}$$

$$-\sigma_{O^{2-}}\Delta\phi / fL = -k''(\mu''_{O_2} - \mu_{O_2}^{air}) = j'' \qquad \text{(A10b)}$$

Here, $\Delta\phi$ is given by

$$V = \Delta\phi + (\mu''_{O_2} - \mu_{O_2}^{c})\big/4e \qquad \text{(A11)}$$

because the Fermi level is constant in the metal. The above equations Eq. (A10-A11) can be solved to obtain $j'$ and $j''$, as well as $\mu'_{O_2} - \mu_{O_2}^{air}$ and $\mu''_{O_2} - \mu_{O_2}^{air}$, for any $V$ and $L$ prescribed,

$$j' = k'(\mu'_{O_2} - \mu_{O_2}^{air}) = (\mu_{O_2}^{c} - \mu_{O_2}^{air})\Big/\Big[1/k' + (1-f)L/k^{metal}\Big] \qquad \text{(A12a)}$$

$$j'' = -k''(\mu'_{O_2} - \mu_{O_2}^{air}) = -\Big[(\mu_{O_2}^{c} - \mu_{O_2}^{air})/4e + V\Big]\Big/\Big[1/4ek'' + fL/\sigma_{O^{2-}}\Big] \text{(A12b)}$$

They are understood as follows: in Eq. (A12a), a total driving force $\mu_{O_2}^{air} - \mu_{O_2}^{c}$ is used to drive $j'$ over the sum resistance of the cathode and metallic suboxide ($fL/4ek^{metal}A$ for $R^{metal}$); in Eq. (A12b), an applied voltage $V$ is used to counter the Nernst potential $-(\mu_{O_2}^{air} - \mu_{O_2}^{c})/4e$ to drive $j''$ over the sum resistance of the anode and YSZ.

While the above solution is correct, $j'$ and $j''$ actually do not match except for $f = 1$ and $V = V^{c}$. The mismatch necessitates a movement of the metal-YSZ interface, across which there is an excess oxygen concentration $\Delta C$. (YSZ, being $0.92ZrO_2+0.08Y_2O_3$, has about 1 more oxygen for every cation than the suboxide, although the exact value of $\Delta C$ depends on the valence of Y in the suboxide.) Such movement creates a convection current, which allows the total oxygen flux to match as follows

$$j' = v(2e\Delta C) + j'' \qquad \text{(A12c)}$$

This gives $v$, which is a function of $f$ and $V$,

$$v(2e\Delta C) = (\mu_{O_2}^{c} - \mu_{O_2}^{air}) \Big/ \left[ 1/k' + (1-f)L/k^{metal} \right] + \left[ (\mu_{O_2}^{c} - \mu_{O_2}^{air}) + 4eV \right] \Big/ \left[ 1/k'' + fL4e/\sigma_{O^{2-}} \right] \quad \text{(A13)}$$

As expected, at $f = 1$, $v = 0$ when $V = V^{c}$, which reduces to the "static" solution obtained in Eq. (A7). When $V > V^{c}$, $v$ is always positive meaning $f$ must decrease with time,

$$dt = -L\,df/v \qquad \text{(A14)}$$

Using $v(f,V)$ of Eq. (A13) and integrating Eq. (A14) on both sides from $t = 0$ and $f = 1$ then provides $f(t)$, and the lifetime $\tau$ is the time it takes for $f$ to reach 0. In the above, the complicated dependence of $v(f,V)$ makes it impossible to realize the simple relation of $\tau \sim L/V$. Lastly, while Eq. (A12c) only satisfies mass continuity for $O^{2-}$, in the cathode and the suboxide there is also an injected electronic current density of $v\Delta C$, which does not require any voltage since these are metallic regions. It does not continue into YSZ because at the metal-YSZ interface, it is completely used to convert $Zr^{4+}$ in YSZ to $Zr^{2+}$ in ZrO. The solution is now complete.

The above solution can be cast into dimensionless form in terms of a characteristic voltage, $V^{*} = (\mu_{O_2}^{air} - \mu_{O_2}^{c})/4e$, and a characteristic time, $t^{*} = (LA2e\Delta C)R^{YSZ0}/V^{*}$, where $R^{YSZ0} = L/A\sigma_{O^{2-}}$ is the resistance of undegraded YSZ of a cross section $A$, and $LA2e\Delta C$ is the total amount of charge required to converting $ZrO_2$ to ZrO

$$-df/d(t/t^{*}) = -1\Big/\left(R^{cathode}/R^{YSZ0} + (1-f)R^{metal}/R^{YSZ0}\right) + \left(V/V^{*} - 1\right)\Big/\left(R^{anode}/R^{YSZ0} + f\right) \quad \text{(A14a)}$$

In the above, $t/t^*$ is the reduced time, $V/V^*$ is the reduced voltage, $R^{\text{cathode}} = 1/4eLk'A$, $R^{\text{anode}} = 1/4eLk''A$, and $R^{\text{metal}} = L/4ek^{\text{metal}}A$ is the (diffusion) resistance of a ZrO (i.e., fully degraded YSZ) of a cross section *A* and a length *L*. Since this equation has the simple form of $-df/d(t/t^*) = -1/\left(a+(1-f)b\right) + \left(V/V^* - 1\right)/\left(c+f\right)$, one can express the critical voltage for getting negative $df/dt$ as $V/V^* = 1 + c/a + 1/a$, which reduces to Eq. (A8). So, Eq. (A14a) can be evaluated to obtain $f(t/t^*)$ for any $V/V^*$ greater than $1 + c/a + 1/a$.

A simplification of the above solution assuming $k''$ and $k^{\text{metal}}$ are infinite is possible under suitable circumstances, but even in such case, *v* is not linearly dependent on $V/L$ only, giving

$$(-L\,df/dt)(2e\Delta C) = k'(\mu_{O_2}^{c} - \mu_{O_2}^{\text{air}}) + (\sigma_{O^{2-}}/fL4e)\left[(\mu_{O_2}^{c} - \mu_{O_2}^{\text{air}}) + 4eV\right] \qquad \text{(A15)}$$

The circumstances are: first, very fast anode kinetics allowing $k''$ to become infinite, and second, the metallic suboxide, which behaves like an electrode, admits oxygen if the side surface of the sample is exposed to air. Indeed, the reduced form of zirconia, sometimes broadly referred to as blackened zirconia, has reportedly good catalytic performance. In such case, we can let $k^{\text{metal}}$ be infinite and replace $k'$ by the reaction constant of the suboxide, denoted as $k'^{\text{metal}}$. It can be shown that, for this problem, the effective $V^{\text{c}}$ decreases with time because of the effective sample length is now $fL$, hence giving a higher field. This will lead to an even faster degradation of resistance, but other main characteristics of the above solution remain unchanged.

**Degradation and reversal**

We already mentioned that $f(t)$ is an indicator of the resistance degradation in the normalized form. If the electrode resistance, which is $1/4eLk'A$ is $R^{\text{cathode}}$ and $1/4eLk''A$ for $R^{\text{anode}}$ according to Eq. (A7), remains unchanged, then the only degradation comes from the resistance of YSZ, denoted by $R^{\text{YSZ}}$. (Strictly speaking, $R^{\text{YSZ}}$ is the small-voltage resistance because at large voltage, YSZ even without considering electrode resistance is not Ohmic, especially at low temperature.) For a sample with a cross sectional area *A*, $R^{YSZ} = fL/A\sigma_{\text{O}^{2-}}$. So, we can equate the initial rate of resistance degradation to that of $R^{\text{YSZ}}$, which is

$$dR^{YSZ}/dt = -v/A\sigma_{\text{O}^{2-}} \qquad \text{(A16)}$$

Noting that at $f=1$, $v$ is proportional to $V-V^{\text{c}}$ when $V>V^{\text{c}}$ in Eq. (A13), we obtain

$$dR^{\text{YSZ}}/dt\Big|_{t=0} = -a\left[(V-V^{\text{c}})/L\right]\Big/\left(1+\sigma_{\text{O}^{2-}}/4eLk''\right) = -a\left[(V-V^{\text{c}})/L\right]\Big/\left(1+R^{\text{anode}}/R^{\text{YSZ}}\right) \quad \text{(A17)}$$

where $a = 1/2e\Delta CA$. This result may be used to estimate the critical field, $V^{\text{c}}/L$. Interestingly, the initial degradation rate is unrelated to the cathode kinetics even though the latter is the origin of degradation that sets $V^{\text{c}}$. Moreover, if $R^{\text{anode}}/R^{\text{YSZ}}$ is small, then the initial degradation rate is unrelated to the cathode electrode kinetics either. This leaves the degradation rate to be independent of almost all the material parameters other except $\Delta C$; it is even independent of the $\sigma_{\text{O}^{2-}}$ of YSZ, thus independent of the temperature and YSZ's composition. This is because that although the conversion of YSZ to metallic suboxide results in a resistance decrement proportional to YSZ's resistivity or $1/\sigma_{\text{O}^{2-}}$, the rate of conversion is governed by $\text{O}^{2-}$ transport in YSZ that is proportional $\sigma_{\text{O}^{2-}}$, and the two factors cancel out. To apply this result, we should note that the resistance is much higher at lower temperatures, so the same degradation rate may seem imperceptible. Another note of caution is that Eq. (A17) applies to the steady state, which is established at $t = 0$, but this may not be the case at low test temperatures.

As already mentioned, the velocity increases with decreasing $fL$ so the degradation rate under a constant $V$ accelerates with time. This is because in Eq. (A12), under the same driving force of constant $V$ and $\mu_{O_2}^{air} - \mu_{O_2}^{c}$, the current density in (a) the cathode-metallic suboxide side decreases with increasing length of the suboxide, whereas in (b) the YSZ-cathode side it increases with decreasing length of YSZ, thus causing an increasing mismatch between $j'$ and $j''$. This prediction can be directly verified by checking the two terms on the right-hand side of Eq. (A13). Doing the same and further dividing both sides by $R^{YSZ} = fL/A\sigma_{O^{2-}}$, we also find

$$\begin{aligned} -(d\ln R^{\text{YSZ}}/dt)(2e\Delta C) = (\mu_{O_2}^{c} - \mu_{O_2}^{air}) \big/ fL\left[1/k' + (1-f)L/k^{\text{metal}}\right] + \\ \left[(\mu_{O_2}^{c} - \mu_{O_2}^{air}) + 4eV\right] \big/ fL\left[1/k'' + fL4e/\sigma_{O^{2-}}\right] \end{aligned} \tag{A18}$$

So the same trend is seen in $-(d\ln R^{\text{YSZ}}/dt)$.

Although we will not attempt a full solution of the transient problem, it is easy to see that, once the metallic suboxide has formed, a decrease in $V$ could cause a pause of the movement of the metal-YSZ interface at a certain $V^0$, and a backward movement at $V < V^0$. Assuming $\mu'_{O_2}$ and $\mu''_{O_2}$ remain the same during the transient, we can obtain $V^0$ by noting (i) on the cathode-suboxide side, $j'$ in Eq. (A10a) driven by $\mu_{O_2}^{air} - \mu_{O_2}^{c}$ will maintain as a positive oxygen flux; (ii) on the anode-YSZ side, $\Delta\phi = V^0 - (\mu''_{O_2} - \mu_{O_2}^{c})/4e$ will drive a positive oxygen flux, $j'' = -\sigma_{O^{2-}}\Delta\phi / L$; and (iii) $j' = j''$ at $V^0$. Thus,

$$V^0 = V\Big/\left[1 + fL4ek''/\sigma_{O^{2-}}\right] - \left[(\mu^{c}_{O_2} - \mu^{air}_{O_2})/4e\right]$$
$$\left[\left(fL4ek''/\sigma_{O^{2-}}\right)\Big/\left(1 + fL4ek''/\sigma_{O^{2-}}\right) + \left(fL4ek'/\sigma_{O^{2-}}\right)\Big/\left(1 + (1-f)Lk'/k^{metal}\right)\right] \quad \text{(A19)}$$

Since usually $V \gg (\mu^{air}_{O_2} - \mu^{c}_{O_2})/4e$, to gain some understanding of Eq. (A19), we may focus on the first term on the right hand side, which states that a smaller voltage drop, namely a $V^0$ closer to *V*, is realized when $fL4ek''/\sigma_{O^{2-}} \ll 1$. This condition corresponds to a larger $\mu''_{O_2}$ at the anode, either to overcome the slower anode kinetics (small $k''$) or to accommodate a faster flux in YSZ (small resistance $fL/A\sigma_{O^{2-}}$), which will lower $\Delta\phi$ that drives the positive oxygen flux in YSZ. Specifically, we expect the interface movement to be more easily reversed by a smaller voltage drop when the remaining length of YSZ is short.

In the above derivation, we assumed $\mu'_{O_2}$ and $\mu''_{O_2}$ remained the same during the transient. This assumption needs justification. At the anode where $\mu_{O^{2-}}$ is constant, $\mu''_{O_2}$ is controlled by $\mu_e$, which can be instantly altered by injecting electrons. (This will also quench oxygen exit.) However, because the electron mobility is very low in YSZ, even lower than that of $O^{2-}$, modifying $\mu_{O_2}$ further inside will take some time. Therefore, the effective $\mu''_{O_2}$ that influences $\Delta\phi$ will remain little changed for a while until $\mu''_{O_2}$ is completely dissipated. At the cathode where the metallic suboxide has formed, $\mu_e$ is constant but $\mu_{O^{2-}}$ is controlled by the concentration of $O^{2-}$ interstitials, which need time to adjust. So the above assumptions seems justified until $\mu'_{O_2}$ and $\mu''_{O_2}$ relax.

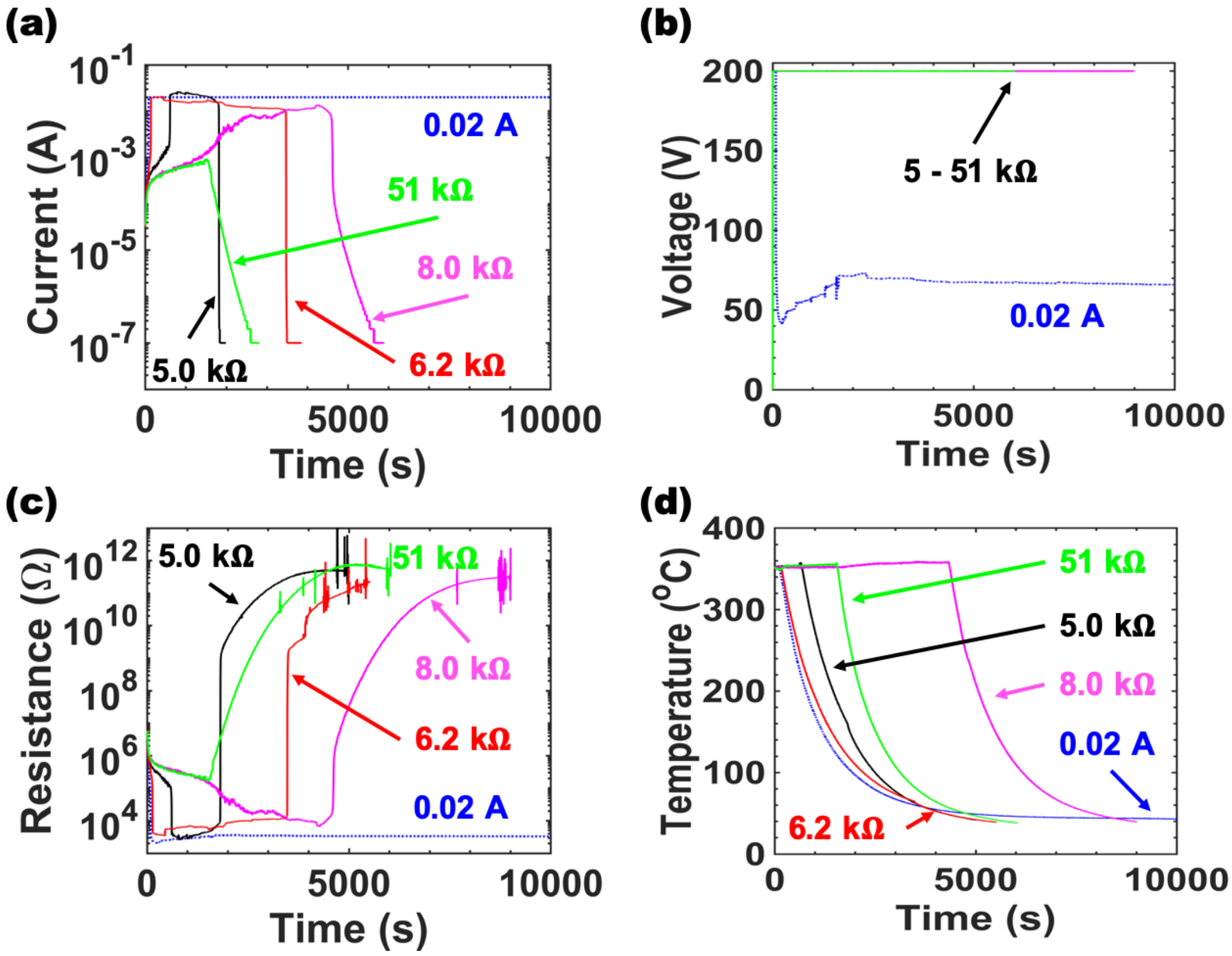

**Fig. S1** Time dependence of the (a) current, (b) voltage, (c) resistance and (d) temperature of a degraded (200 V at 350°C) single crystal cooled after resistance reaching $2R_L$, hence with a flash at $R_L < 7$ kΩ or without a flash at $R_L > 7$ kΩ. Also shown is the dependence when cooled after current reaching a compliance limit of 0.02 A, which was maintained during cooling.

**Fig. S2** Video, at normal speed, of a polycrystalline 8YSZ near flash after being tested at 335ºC and 200 V on a hot plate. The cathode is at the bottom and the anode at the top. After a few seconds of flashing, some of the defects are annealed away and the sample turns back to white. Need to Acrobat Reader to see video.

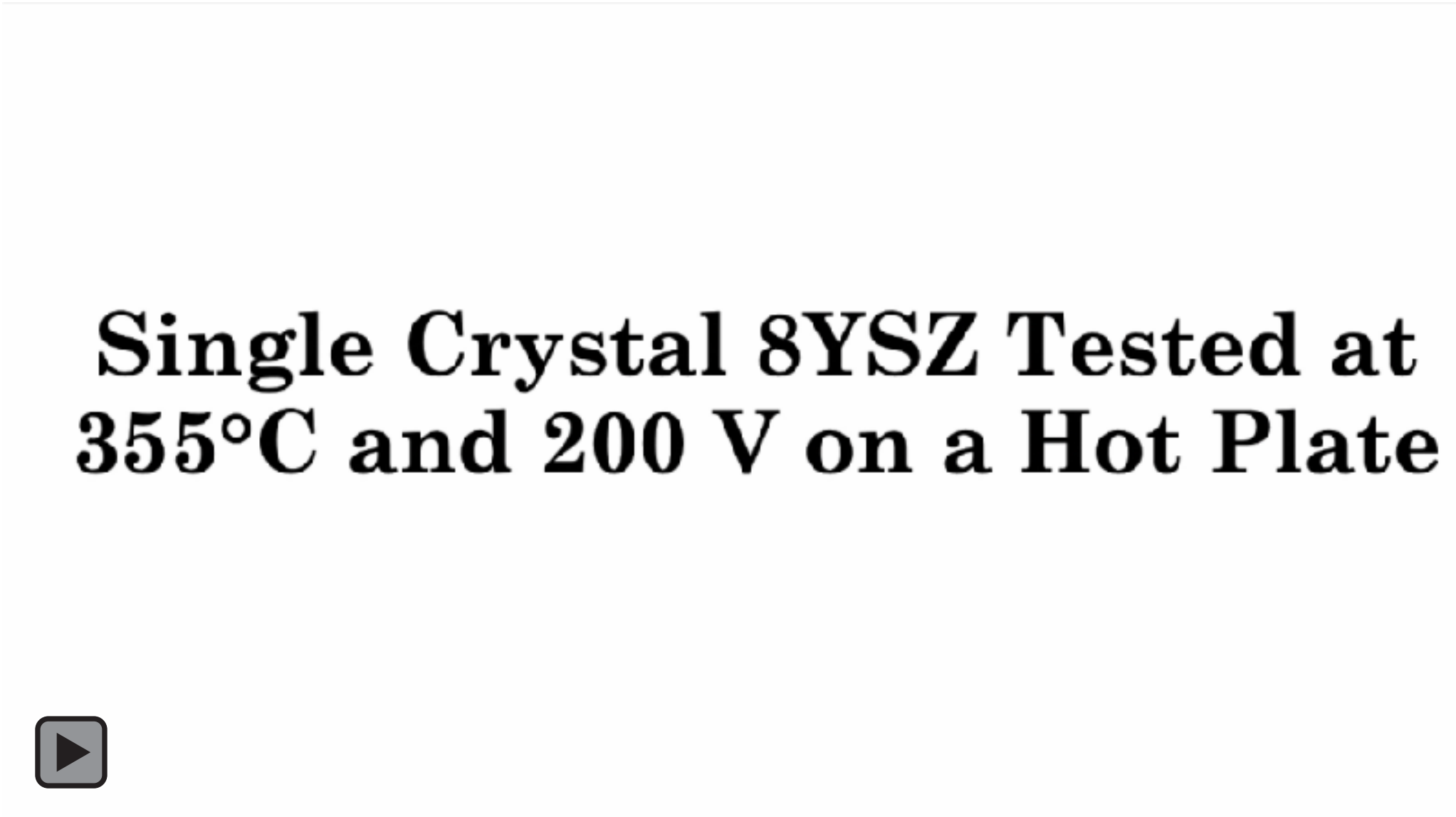

**Fig. S3** Full video, at normal speed, of a single crystal 8YSZ tested at 355ºC and 200 V on a hot plate. The cathode is at the bottom and the anode at the top. The flash begins at the anode and the sample ends up breaking. Need Acrobat Reader to see video.

(a)

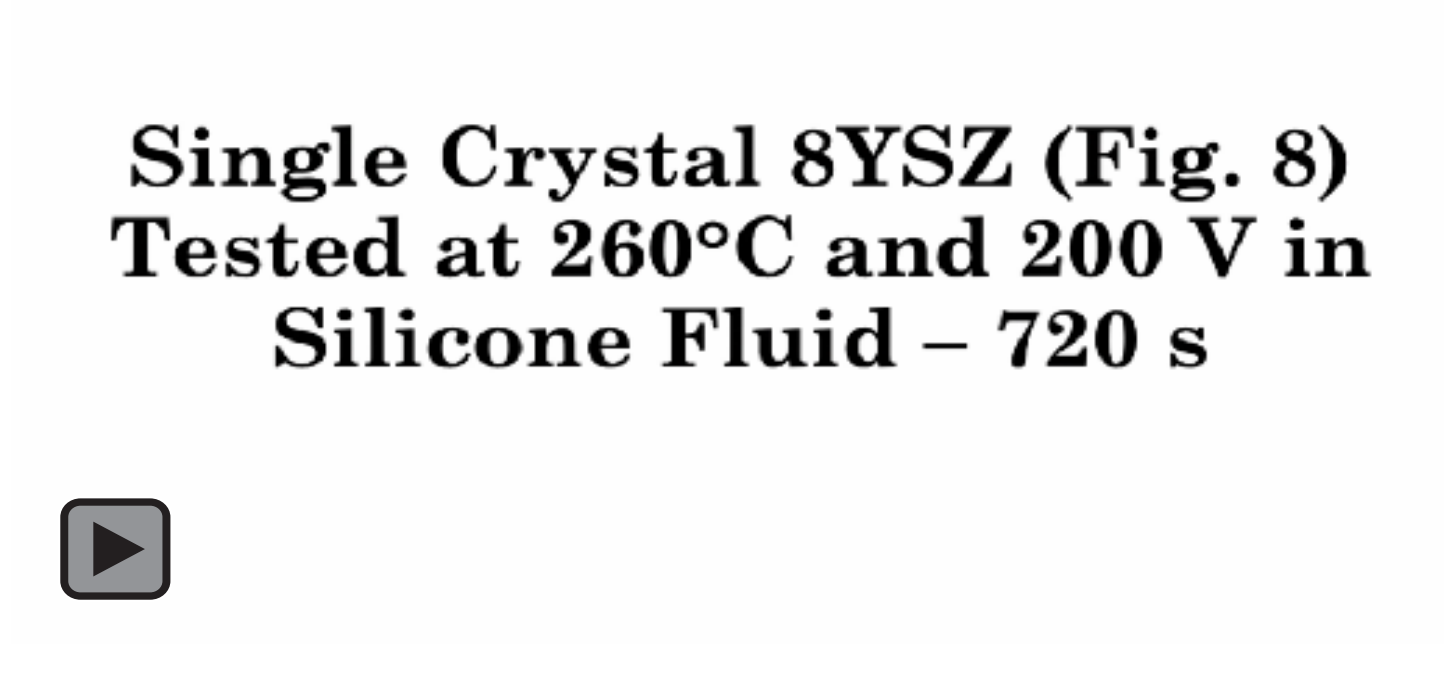

(b)

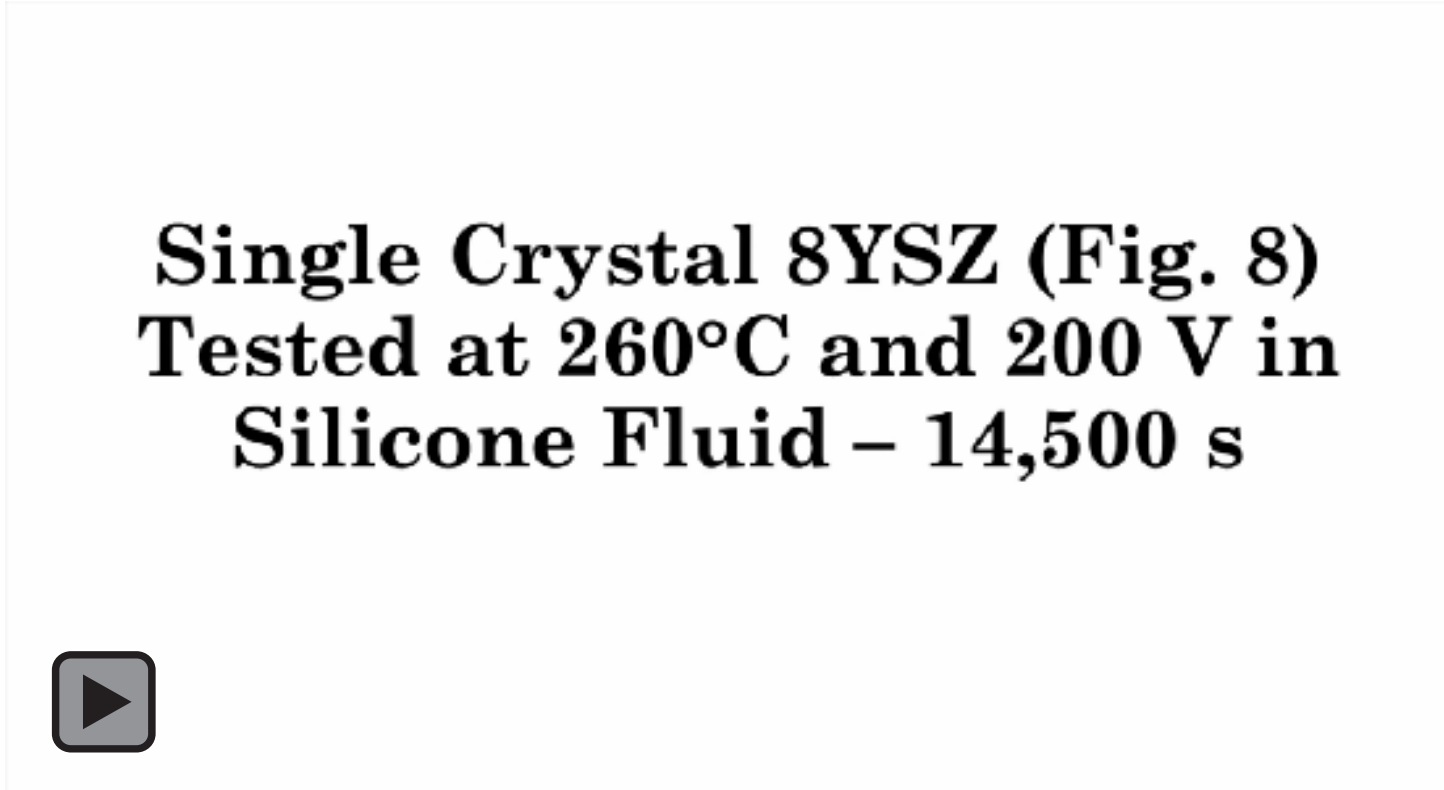

(c)

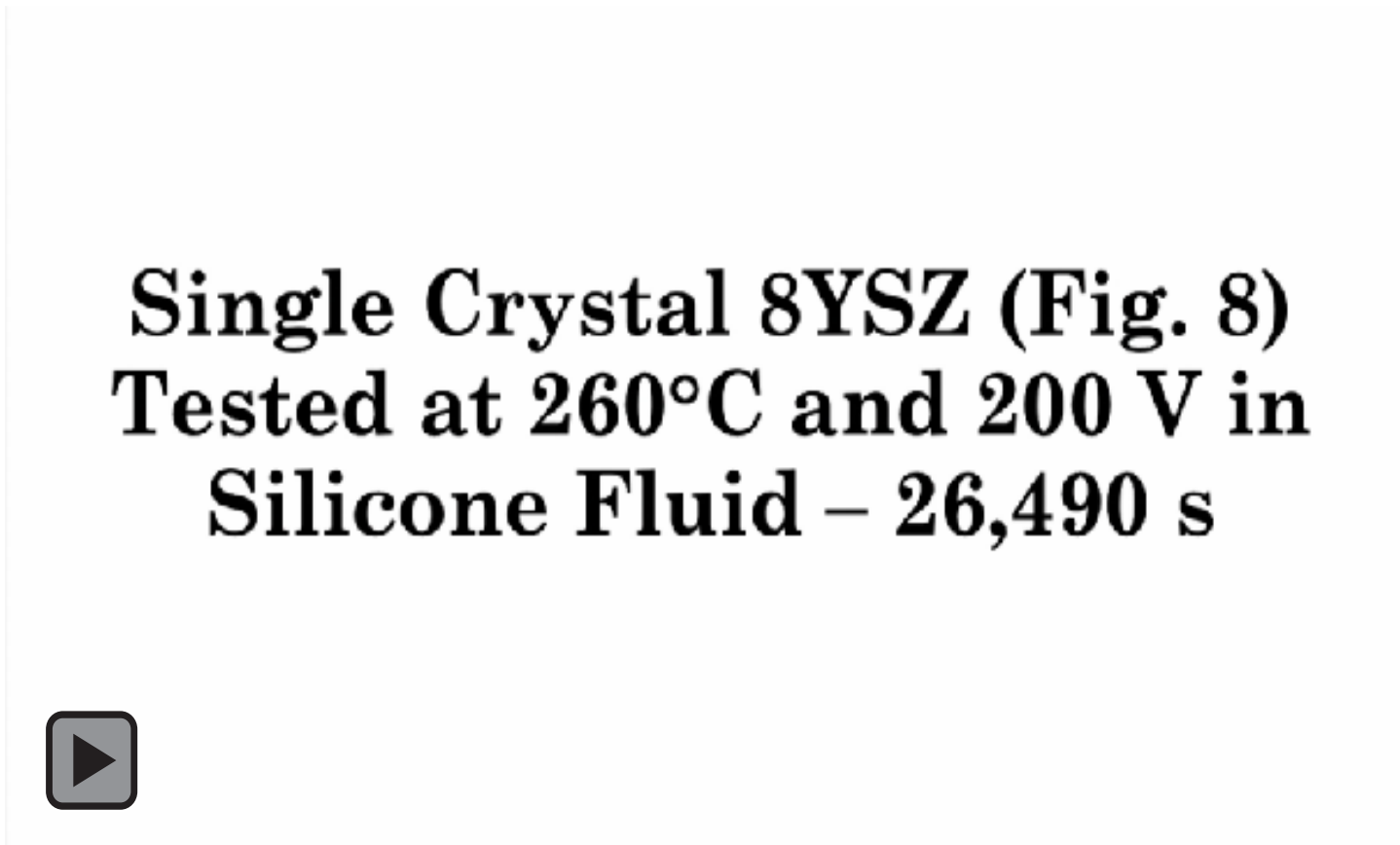

**Fig. S4** Video, at normal speed, of the single crystal found in **Fig. 8** in the main text, tested at 260ºC under 200 V in silicone fluid recorded at (a) t = 720 s, (b) t = 14,500 s and (c) t = 26,490 s. The cathode is on the left and the anode on the right. Although bubbles all originated from the anode, they often were attached to the sample edge, growing and migrating until becoming big enough to detach. Need to Acrobat Reader to see video.